\documentclass{aa}  

\usepackage{graphicx}
\usepackage{txfonts}
\usepackage{xcolor}
\usepackage{array,multirow,makecell} 
\usepackage[breaklinks, colorlinks, citecolor=blue]{hyperref}
\usepackage{enumitem}
\def\drm{\mathrm{d}}
\def\los{{\bf \hat n}}
\def\vvec{{\bf v}}
\def\zre{{z_\mathrm{re}}}
\def\zend{{z_\mathrm{end}}}

\newcommand{\Planck}{{\it Planck}}
\def\DkSZ{\mathcal{D}^{\rm kSZ}_{3000}}
\def\DtSZ{\mathcal{D}^{\rm tSZ}_{3000}}

\defcitealias{PAPER1}{Paper I}
\defcitealias{PAPER3}{Paper III}
\defcitealias{Reichardt2020}{R21}

\begin{document}

   \title{Retrieving cosmological information from small-scale \\CMB foregrounds}

   \subtitle{II. The kinetic Sunyaev Zel'dovich effect}

   \author{Adélie Gorce\inst{1}
          \and
          Marian Douspis\inst{2}
          \and 
          Laura Salvati\inst{2}
          }

   \institute{
   Department of Physics and McGill Space Institute, McGill University, Montreal, QC, Canada H3A 2T8 
   \and
    Université Paris-Saclay, CNRS,  
    Institut d'Astrophysique Spatiale, 
    91405, Orsay, France    \\
    \email{adelie.gorce@mcgill.ca}
             }
\date{Received 17 February 2022 / Accepted 22 April 2022}

 
  \abstract{
  Recent results of ground-based telescopes, giving high-quality measurements of the cosmic microwave background (CMB) temperature power spectrum on small scales (below $ 1\,\mathrm{deg}$) motivate the need for an accurate model of foregrounds, which dominate the primary signal at these multipoles.
  In a previous work, we have shown that cosmological information could be retrieved from the power spectrum of the thermal Sunyaev Zel'dovich (SZ) effect. In this work, we introduce a physically motivated model of the Epoch of Reionisation in the cosmological analysis of CMB data, which is coherent on all scales. In particular, at high multipoles, the power spectrum of the kinetic SZ (kSZ) effect is inferred from a set of cosmological and reionisation parameters by a machine-learning algorithm.

  First including an asymmetric parameterisation of the reionisation history in the \Planck~2018 data analysis, we retrieve a value of the Thomson optical depth consistent with previous results, but stemming from a completely different history of reionisation in which the first luminous sources light up as early as $z=15$. Considering the latest small-scale data from the South Pole telescope (SPT) and letting the cosmology free to vary, we find that including the new cosmology-dependent tSZ and kSZ spectra helps tighten the constraints on their amplitudes by breaking their degeneracy. We report a $5\sigma$ measurement of the kSZ signal at $\ell=3000$, $\mathcal{D}_{3000}^\mathrm{kSZ} = 3.4^{+0.5}_{-0.3}\,\mu\mathrm{K}^2$ at the 68\% confidence level (C.L.), marginalised over cosmology, as well as an upper limit on the patchy signal from reionisation $\mathcal{D}_{3000}^\mathrm{pkSZ}<1.6~\mu\mathrm{K}^2$ (95\% C.L.). Additionally, we find that the SPT data favour slightly earlier reionisation scenarios than \Planck, leading to $\tau = 0.062 ^{+0.012}_{-0.015}$ and a reionisation midpoint $\zre = 7.9^{+1.1}_{-1.3}$ (68\% C.L.), which is in line with constraints from high-redshift quasars and galaxies.
   }

\maketitle
%

\section{Introduction}\label{sec:intro}

Different generations of satellites, from the cosmic background explorer \citep[COBE,][]{COBE} to, most recently, \Planck~\citep{Planck2018}, led us to exquisite measurements of the power spectrum of the large-scale fluctuations of the cosmic microwave background (CMB) temperature. Today, efforts are turned towards ground-based observatories of such fluctuations on small scales. The Atacama cosmology telescope \citep[ACT,][]{ACT}\footnote{\url{https://act.princeton.edu}} and the South Pole telescope \citep[SPT,][]{SPT}\footnote{\url{http://pole.uchicago.edu}} target this signal, with the latter recently publishing observations of sufficient quality \citep[][hereafter R21]{Reichardt2020} to allow for the first $\geq 3\sigma$ measurement of the amplitude of the angular power spectra of the kinetic and thermal Sunyaev-Zel'dovich effects \citep[kSZ and tSZ, respectively,][]{zeldovich_sunyaev_1969,sunyaev_zeldovich_1980} at an angular multipole of $\ell=3000$.

Along with thermal dust emission from dusty star-forming galaxies, radio galaxy emission, the Galactic cirrus signal, and the cosmic infrared background (CIB), the tSZ and kSZ power spectra are foregrounds to the CMB primary anisotropies. That is, their power is not sourced by temperature fluctuations in the primordial plasma at the time of recombination, but from the interaction of the photons released at the surface of last scattering with astrophysical objects along the line of sight, before reaching the detector. In this respect, and despite the fact that they largely dominate the primary signal on scales smaller than one arcminute, their amplitudes have often been considered in analyses of CMB data as nuisance parameters \citep[e.g.][]{reichardt_2012_spt,DunkleyCalabrese_2013,Ichiki_2014,GeorgeReichardt_2015}. A common approach consists of fitting a primary power spectrum derived from sampled cosmological parameters and a sum of foregrounds power spectra modelled by templates to the observed data points. Only the amplitudes of the various foregrounds, at a given multipole (often, $\ell=3000$), are fitted to data.

However, these foregrounds trace the matter the CMB photons encounter on their way to us and they include cosmological information which cannot be retrieved by the use of templates \citep{2006A&A...456..819D,ShawNagai_2010, Salvati2018, ManiyarLagache_2019, Gorce2020}. In a first work \citep[][hereafter Paper I]{PAPER1}, we have shown that replacing the template used for the tSZ power spectrum by a cosmology-dependent spectrum when analysing data from \Planck~and the SPT tightens constraints on cosmological parameters. In this work, we perform a similar analysis, this time focusing on constraints obtained when including a cosmology-dependent kSZ angular power spectrum in the analysis.\\

The kSZ power spectrum traces the energy gained by CMB photons interacting with electrons that have a non-zero bulk velocity relative to the CMB rest-frame along the line-of-sight.
Numerous works \citep[e.g.][]{mcquinn_2005,iliev_2007,mesinger_2012_kSZ,ShawRudd_2012,battaglia_2013_paperIII,Alvarez_2016, ChenTrac_2022} have shown that the amplitude of the kSZ angular power spectrum can be related to the nature of the Epoch of Reionisation (EoR), when the first light sources re-ionised the intergalactic medium (IGM), mostly neutral since recombination and the emission of the CMB photons. Indeed, as an integrated effect, a longer or earlier reionisation process will lead to a larger kSZ amplitude. 
Traditionally, the kSZ power is divided into two contributions: The late-time kSZ accounts for CMB photons scattering off electrons once reionisation is over and the IGM is fully ionised, whilst the patchy kSZ accounts for photons scattering off ionised bubbles along the otherwise neutral line of sight during reionisation. In contrast to the late-time kSZ, not only the amplitude, but also the shape of the patchy kSZ power spectrum can tell us about the reionisation process \citep{mcquinn_2005, iliev_2007, mesinger_2012_kSZ}. For example, the multipole at which the patchy kSZ reaches its maximum has been related to the characteristic size of ionised bubbles during the EoR, in turn related to the physical properties of early galaxies, if those are assumed to have led the reionisation process \citep{Bouwens2015,Robertson2015,GreigMesinger_2017,GorceDouspis_2018}.

With the new generation of the SPT \citep{SPT3G}, the Simons Observatory \citep{Ade_2019} and, later, CMB-Stage~4 experiments \citep[CMB-S4,][]{CMB-S4}, coming online, there is a growing interest in the community for the potential of small-scale CMB data to constrain reionisation. A common approach, popularised by the sampler of 21cmFAST simulations \citep{21cmFAST_2011,  ParkMesinger_2019, Murray2020}, 21CMMC \citep{GreigMesinger_2015}, is to sample astrophysical parameters in semi-numerical simulations and confront the observables derived at each iteration to data \citep{QinPoulin_2020,ChoudhuryMukherjee_2021}. However, such approaches are limited by computational power, both in terms of the precision and the size of the employed simulations. Very large box sizes are necessary to capture the large-scale velocity flows contributing to the kSZ effect at high-$\ell$ and results based on insufficiently large simulations will significantly underestimate its power \citep{ShawRudd_2012}.

To circumvent these numerical difficulties, we have introduced  in \citet{Gorce2020} a new analytical derivation of the kSZ angular power spectrum, based on the cosmology and the reionisation history. With this formalism, it is possible to compute the theoretical kSZ power spectrum associated with a set of cosmological and reionisation parameters at each iteration of the sampler, when analysing small-scale fluctuations of the CMB temperature. Constraints on the cosmological model and the EoR are retrieved from the measured shape and amplitude of the kSZ signal. Such an analysis, applied to the latest \Planck~and SPT data \citep{Planck2018,Reichardt2020}, is the focus of the present work. In an approach similar to \citetalias{PAPER1}, we use machine learning to efficiently predict the otherwise computationally intensive derivation of the kSZ power spectrum on the multipole range covered by the data. In order for the reionisation modelling to be consistent across all scales probed, we substitute the commonly used `instantaneous' parameterisation of the evolution of the IGM ionised fraction with redshift \citep{camb1} by a physically motivated asymmetric parameterisation \citep{Douspis2015}, also used in the kSZ derivation. We subsequently assess the impact of this substitution on the cosmological constraints obtained with large-scale data.\\

This paper is organised as follows. In Sec.~\ref{sec:methods}, we detail the theoretical framework used to derive the kSZ power spectrum, and in particular, introduce the asymmetric parameterisation of reionisation used throughout. We also introduce the statistical methods used to analyse the data listed in Sec.~\ref{sec:data}. In Sec.~\ref{sec:results}, we first present the results of applying the asymmetric reionisation model to \Planck~data. We then include the new kSZ modelling in the analysis of the small-scale SPT data and retrieve reliable constraints on the midpoint and duration of reionisation. Finally, we discuss our results in Sec.~\ref{sec:discussion} and conclude in Sec.~\ref{sec:conclusions}.

\section{Methods}
 \label{sec:methods}
 
 \subsection{Asymmetric reionisation history} \label{subsec:1_xe}

In the fiducial analysis of \Planck~large-scale data, the reionisation 
history is not constrained directly. Instead, it is the Thomson optical depth $\tau$ that is sampled, as it is strongly correlated to the amplitude of the high-$\ell$ temperature and low-$\ell$ polarisation CMB power spectra \citep{Planck_2016_reio}. As a measure of the fraction of CMB photons absorbed by free electrons along the line-of-sight, the optical depth is directly related to the ionisation level of the IGM or number of electrons per Hydrogen nucleus $x_e(z)$ \citep{camb1,camb2}:
\begin{equation}
\label{eq:def_tau}
\tau(z) = c\,\sigma_\mathrm{T} \int_{0}^{z} \drm z'\ x_e(z')\, n_b(z')\, \frac{(1+z')^2}{H(z')} ,
\end{equation}
where $\sigma_T$ is the Thomson cross-section, $n_b$ is the average density of baryons and $H$ is the Hubble parameter. Here, for simplicity, the product of $x_e$ and $n_b$, both averaged over the sky, is taken as a proxy for the mean electron density, $\langle x_e \times n_b \rangle$, overlooking the correlation between the electron and the density fields. This can lead to a $5$ to $10\%$ error on the measured value of $\tau$ \citep{LiuPritchard_2016}.
To interpret $\tau$ in terms of the reionisation history, the following symmetric, `instantaneous', parameterisation is often considered:
\begin{equation}
\label{eq:xe_tanh}
x_e(z) = \frac{f_\mathrm{e}}{2}\, \left[ 1 + \text{tanh}\left( \frac{y-y_\mathrm{re}}{\delta y}\right) \right],  
\end{equation}
where $y\, (z)\equiv(1+z)^{\frac{3}{2}}$, $y_\mathrm{re}=y\, (z=z_\mathrm{re})$ for $z_\mathrm{re}$ the midpoint of Hydrogen reionisation, such that $x_e(\zre)=0.5$ and $\delta y \equiv \frac{3}{2}\, (1+z)^{\frac{1}{2}}\, \delta z$ for $\delta z = 0.5$, roughly corresponding to the redshift range which sees the IGM ionised fraction increasing from $25\%$ to $75\%$. The $f_e$ parameter corresponds to the number of electrons per Hydrogen atom, and is equal to $f_e\simeq 1.08$ when accounting for the first ionisation of Helium. The contribution of the second Helium ionisation is added for $z>4.5$ as another hyperbolic tangent such as the one given in Eq.~\eqref{eq:xe_tanh}, with parameters $z_\mathrm{re,He}=5.0$ and $\delta z_{\rm He} = 0.5$ \citep{kuhlen_2012}. It is finished by $z=3.5$.
When $\delta z$ is fixed, it is possible to deduce the value of $\zre$ through binary search from a given $\tau$ value. 

In this framework, the evolution of $x_e$ with redshift is a quick and symmetrical process, centred on $\zre$.
However, recent simulations and analyses of astronomical data sets show the reionisation process to be a highly gradual evolution, starting slowly at redshifts as high as $z=15$ \citep{Robertson2015,GorceDouspis_2018}. Since the kSZ power depends largely on the duration of reionisation \citep[see Sec.~\ref{subsec:1_kSZ} and][]{battaglia_2013_paperIII,Alvarez_2016} and on its asymmetry \citep{ParkShapiro_2013,ChenTrac_2022}, it is essential to include a more accurate evolution of $x_e(z)$ in any analysis attempting to measure its amplitude. Accounting for this asymmetry can be done by, for instance, interpolating between different ionisation levels \citep[e.g.][]{MilleaBouchet_2018, TracChen_2022} or by changing the parameterisation for the redshift-evolution of the ionisation level. In this work, we follow this second approach and use the power-law introduced in \citet{Douspis2015} and already confronted to large-scale CMB data in \citet{Planck_2016_reio}:
\begin{equation}
\label{eq:xe_param}
x_e(z)  = 
\left\{ 
	\begin{array}{ll}
		f_\mathrm{e} & \mathrm{for} \: z< z_\mathrm{end},\\
		f_\mathrm{e}\, \left(\frac{z_\mathrm{early}-z}{z_\mathrm{early}-z_\mathrm{end}}\right)^\alpha & \mathrm{for} \: z>z_\mathrm{end},
	\end{array}
\right.
\end{equation}
where $z_\mathrm{early}$ corresponds to the redshift around which the first emitting sources form and at which $x_e(z)$ is matched to a residual ionised fraction  of $10^{-4}$. We fix $z_\mathrm{early}=20$, in agreement with current constraints on the history of reionisation obtained from astrophysical data sets \citep{Robertson2015,GreigMesinger_2015,GorceDouspis_2018}. Additionally, we find that the data considered in this work does not require sources to light up before $z=20$. We let $z_\mathrm{end}$ and $z_\mathrm{re}$ vary, since fixing $z_\mathrm{early}$ makes $\zre$ a sampling equivalent to $\alpha$ (Eq.~\ref{eq:xe_param}), and we choose the former to facilitate the interpretation of results. Finally, for convergence purposes, we sample $\mathrm{d}z \equiv \zre-\zend$ instead of $\zend$ and we consider the contribution of Helium reionisation. We include this parameterisation into the derivation of the kSZ angular power spectrum detailed below.

\subsection{Derivation of the kSZ power spectrum} \label{subsec:1_kSZ}
 
The temperature anisotropies sourced by the scattering of CMB photons off clouds of free electrons with a non-zero bulk velocity $\vvec$ relative to the CMB rest-frame along the line of sight $\los$ write
\begin{equation}
\label{eq:delta_TkSZ}
  \delta T_{\rm kSZ}(\los) =\frac{\sigma_T}{c} \int \frac{\drm \eta}{\drm z}\frac{\drm z}{(1+z)}\, \mathrm{e}^{-\tau(z)}\, n_e(z) \, \vvec\cdot\los \;,
\end{equation}
with $\eta$ the comoving distance to redshift $z$ and $\vvec\cdot\los$ the component of the peculiar velocity of the electrons along the line of sight. The corresponding angular power spectrum writes $C_\ell^{k \rm SZ} \equiv T^2_\mathrm{CMB} \vert \hat{\delta T}_\mathrm{kSZ}(k)\vert^2$, where $T_\mathrm{CMB}$ is the mean temperature of the CMB and $\hat{\delta T}_\mathrm{kSZ}$ is the Fourier transform of $\delta T_{\rm kSZ}$ given above, such that $k = \ell/\eta$ for $\ell$ the angular multipole. In the following, we present results in terms of $\mathcal{D}_\ell \equiv \ell (\ell +1 )C_\ell /2\pi$. 

As already mentioned in Sec.~\ref{sec:intro}, the kSZ signal is often decomposed into a late-time and a patchy component. In \citet{Gorce2020}, we introduced an analytical derivation of the kSZ angular power spectrum based on the reionisation history $x_e(z)$ and two parameters related to the morphology of reionisation. Firstly, the drop-off frequency of the power spectrum of free electrons density fluctuations, $\kappa$, which is directly related to the angular scale where the patchy kSZ power reaches its maximum, and, in turn, to the typical size of ionised bubbles during the EoR. It is expressed in $\mathrm{Mpc}^{-1}$. Secondly, the large-scale amplitude of the electron power spectrum, $\alpha_0$, which is a representation of the variance of the electron field and, as such, is related to the morphology of the ionisation field. It is expressed in $\mathrm{Mpc}^3$.

These two parameters are directly related to the shape and amplitude of the patchy component of the kSZ spectrum. The late-time contribution is derived using a biased matter power spectrum to describe the low-redshift spatial distribution of free electrons, according to the model introduced in \citet{ShawRudd_2012}, fitted to highly resolved simulations \citep{aubert_2015_EMMA} -- see \citet{Gorce2020} for more details on the model.
Instead of using a template of the kSZ power spectrum and fitting its amplitude to small-scale CMB data, in this work, we sample $\alpha_0$ and $\kappa$, along with $\zre$ and $\zend$, so that the kSZ spectrum can be fully computed at each iteration of the sampler. In general, we expect the modelling uncertainties of the parameterisation, including the additional kSZ power stemming from the non-Gaussianity of the electron density field \citep{Alvarez_2016} to be absorbed by the uncertainties on the sampled parameters. However, once observations and foreground-modelling allow for a detailed detection of the kSZ spectra, a more precise description will be required.

\begin{figure}
   \centering
   \includegraphics[width=\columnwidth]{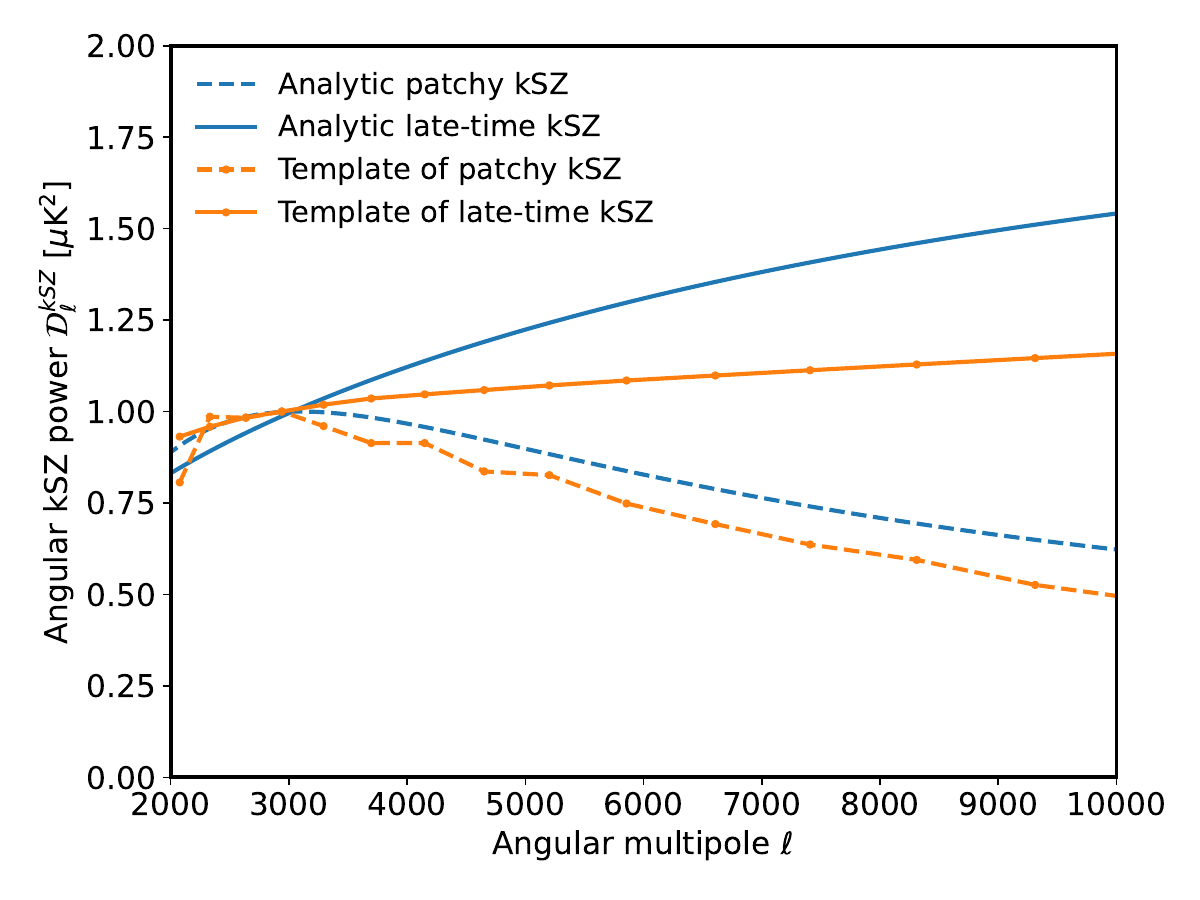}
      \caption{Comparison of the normalised kSZ angular power spectra obtained for a set of reionisation parameters using binned templates \citep{Reichardt2020}, in orange, and using the analytic derivation introduced in \citet{Gorce2020}, in blue.}
    \label{fig:ksz_template_vs_analytic}
   \end{figure}

In Fig.~\ref{fig:ksz_template_vs_analytic}, we show the difference in the late-time and patchy kSZ spectra corresponding to the templates used in \citetalias{Reichardt2020}, and obtained using the analytic derivation of \citet{Gorce2020} for parameters fitted on a semi-numerical simulation \citep{rsage}. In order to highlight shape differences, the spectra are normalised to $1\,\mu\mathrm{K}^2$ at $\ell=3000$. The template used for the late-time signal is extracted from the work of \citet{ShawRudd_2012}, who fit the power spectrum of electron number density fluctuations, considered as a biased matter power spectrum, on hydrodynamic simulations. The template used for the patchy signal was first introduced in \citet{zahn_2012_spt} and was computed numerically from a full radiative transfer simulation \citep{ZahnLidz_2007,ZahnMesinger_2011}. There is a clear difference in the shapes obtained in both cases, in particular on small scales. This is related to the fact that the low-redshift section of our analytic kSZ model was calibrated on a simulation which uses adaptive mesh refinement, and, hence, resolves high-density regions very well, making the derived spectra more sensitive to the thermal behaviour of the gas, and boosting power on small scales \citep{aubert_2015_EMMA}. The results presented in Sec.~\ref{sec:results} are, however, robust to these modelling uncertainties.%

The cosmology-dependent derivation of the kSZ power spectrum is computationally expensive because of the complexity of Eq.~\eqref{eq:delta_TkSZ}, which includes several integrals. Additionally, the time required to compute a $C_\ell$ increases with $\ell$, making the analysis of small-scale data challenging. 
For this reason, it cannot be included as such in a parameter sampling algorithm. Similarly to what was proposed in \citetalias{PAPER1}, we choose to solve this issue by using a machine learning algorithm, random forests (RF), trained on a set of kSZ spectra computed according to \citet{Gorce2020}, to predict the $C_\ell$'s at each iteration of the sampler. The training set is made of 50~000 different spectra, computed on a range of cosmological and reionisation parameters, for 25 angular multipoles linearly distributed over the range $100 \leq \ell \leq 10\,500$, which includes the SPT data points. The details of the RF training and testing can be found in Appendix~\ref{app:ML}. Namely, the errors coming from the RF reconstruction never exceed $0.02~\mu\mathrm{K}^2$, about fifty times smaller than the precision of current measurements \citep{Reichardt2020}. The overall scores of the patchy and late-time predictions are, respectively, $99.3\%$ and $99.8\%$. The RF prediction code for the kSZ power spectra is publicly available\footnote{See \url{https://szdb.osups.universite-paris-saclay.fr/} for a user-friendly interface and a link to the public repository.}.

\subsection{MCMC analysis}

We use a version of the publicly available Monte-Carlo Markov Chain sampler \texttt{CosmoMC} \citep{lewis_2002_cosmomc,cosmosmc2}, modified to include the SPT likelihood, available online \citep{Reichardt2020}, and the \Planck21-tSZ likelihoods, based on \citet{TanimuraDouspis_2021} data and following \citet{Salvati2018}. We further modify the sampler to allow for an on-the-fly RF-prediction of the tSZ, patchy kSZ and late-time kSZ angular power spectra (see above and App.~\ref{app:ML}), as initiated in \citetalias{PAPER1}. 
For the derivation of the tSZ power spectrum, we follow the recipe detailed in \citetalias{PAPER1}. We add in the model the scaling relation between the tSZ signal (the Compton parameter $Y_{500}$) and the halo mass $M_{500}$, both defined within a radius that contains 500 times the critical density of the Universe. In this analysis, for the scaling relation we include as a varying parameter only the mass bias $(1-b)$, defined as the ratio between the halo mass estimated from the hydrostatic equilibrium assumption $M_{\text{HE}}$, and the total halo mass, that is $(1-b) = M_{\text{HE}}/M_{500}$. We focus on this parameter, since it shows the strongest degeneracy with the cosmological ones. We refer the reader to the full discussion in \citetalias{PAPER1}.
For the derivation of the kSZ power spectrum, we add as varying parameters in the analysis the two quantities introduced in Sec.~\ref{subsec:1_kSZ}, that is $\log \alpha_0$ and $\kappa$. These replace the amplitude of the total kSZ power at $\ell=3000$, $\mathcal{D}_{3000}^\mathrm{kSZ}$, which is now derived.

We modify the \texttt{CAMB} module \citep{camb1,camb2}, included within \texttt{CosmoMC} to perform theoretical computations, to account for the asymmetric parameterisation of the reionisation history given in Eq.~\eqref{eq:xe_param}. Doing so, the optical depth $\tau$ is no longer sampled, but derived from the values of the reionisation mid- and endpoint, $\zre$ and $\zend$, themselves sampled. We impose a prior $\zend>4.5$. When the cosmology is let free to vary, we additionally sample the following parameters: 
\begin{itemize}[nolistsep]
    \item The baryon density multiplied by the reduced Hubble constant $h\equiv H_0/100$, $\Omega_b h^2$; 
    \item The cold dark matter density $\Omega_c h^2$; 
    \item The ratio of the sound horizon to the angular diameter distance at decoupling, scaled by 100, $100\theta_\mathrm{MC}$; 
    \item The initial super-horizon amplitude of curvature perturbations at $k = 0.05~\mathrm{Mpc}^{-1}$, $A_s$, as $\log A \equiv \ln (10^{10}A_s)$, and
    \item The scalar spectral index $n_s$.
\end{itemize} 
Finally, nuisance parameters are added depending on the data set used, as described below.

The convergence of the runs is assessed by examining the Gelman-Rubin convergence parameter $R$ \citep{GelmanRubin_1992,AnBrooksGelman_1998}, which corresponds to the ratio of the variance of chain means to the mean of chain variances, for each sampled parameter. We consider the chains converged once, for each parameter, $R-1<10^{-2}$. Unless specified otherwise, all parameter values are given in terms of the maximum of the marginalised posterior probability, and confidence intervals correspond to the highest-density probability intervals at $68\%$\footnote{In contrast to equal-tail probability intervals, the highest-density probability interval is more appropriate for Bayesian analysis with skewed distributions, as it ensures all the values included in the confidence interval have a probability higher than $ 0.68$.}.



\section{Data}\label{sec:data}

\subsection{Planck low-$\ell$ data}\label{subsec:data_planck}

For the analysis of large-scale data presented in Sec.~\ref{subsec:res_planck_asym}, we consider the \citet{Planck2018} data, available in the Planck Legacy archive\footnote{See \url{http://pla.esac.esa.int/pla/}.}. We use the low-$\ell$ $TT$ and $EE$ (\texttt{simall}) likelihoods, as well as high-$\ell$ $TT$, $TE$ and $EE$ spectra (\texttt{plik}) and lensing. The cosmological constraints resulting from the fiducial analysis performed by the \Planck~collaboration are the following (with a $68\%$ confidence): $\Omega_b h^2=0.02237 \pm 0.00015$,  $\Omega_c h^2=0.1200 \pm 0.0012$, $100\theta_\mathrm{MC}=1.04092 \pm 0.00031$, $\tau=0.0544 \pm 0.0073$, $\log A=3.044 \pm 0.014$ and $n_s=0.9649 \pm 0.0042$ \citep{Planck2018_cosmo}.

\subsection{SPT high-$\ell$ data}\label{subsec:data_spt}

We consider the small-scale measurements of the CMB temperature power spectrum presented by the South Pole telescope collaboration in \citet{Reichardt2020} and available online\footnote{Bandpowers and likelihood codes compatible with \texttt{CosmoMC} can be downloaded at \url{https://pole.uchicago.edu/public/data/reichardt20/}.}. The data are made of bandpowers at 95, 150 and 220\,GHz measured over the angular multipole range $2~000 < \ell < 13~500$. The likelihood makes use of auto- and cross-spectra and marginalises over calibration and beam parameters. The total measured temperature power spectrum is considered to be the sum of the primary signal, as well as the following foregrounds: the thermal and kinetic Sunyaev-Zel'dovich effects, Poisson-distributed radio galaxies, whose power goes as $\ell^2$, and the galactic cirrus power. The contribution from thermal dust emission coming from the faint dusty star forming galaxies that cannot be masked out, and forming the CIB \citep{LagachePuget_2005,CaseyNarayanan_2014}, is also considered. The power is divided between spatially clustered and Poisson-distributed galaxies \citep{VieroWang_2013}, with the latter following $\mathcal{D}_\ell \propto \ell^2$.
These foregrounds, in the initial analysis, are all modelled with templates, whose amplitude at $\ell=3000$ is adjusted to fit the data. To this list is added the (negative) cross-spectrum of the CIB and the tSZ signal \citep{AddisonDunkley_2012} and derived as a combination of the \citet{ShangHaiman_2012} model and the tSZ $C_\ell$'s. The absolute value of the amplitude at $\ell=3000$ of the cross-spectrum, labelled $\xi$, is sampled. This results in a list of eight nuisance parameters (Fig.~\ref{fig:app_nuisance_reichardt_comparison}).

\section{Results}
\label{sec:results}

\subsection{Substituting the reionisation model in low-$\ell$ data}
\label{subsec:res_planck_asym}

 \begin{figure*}
    \centering
    \includegraphics[width=\textwidth]{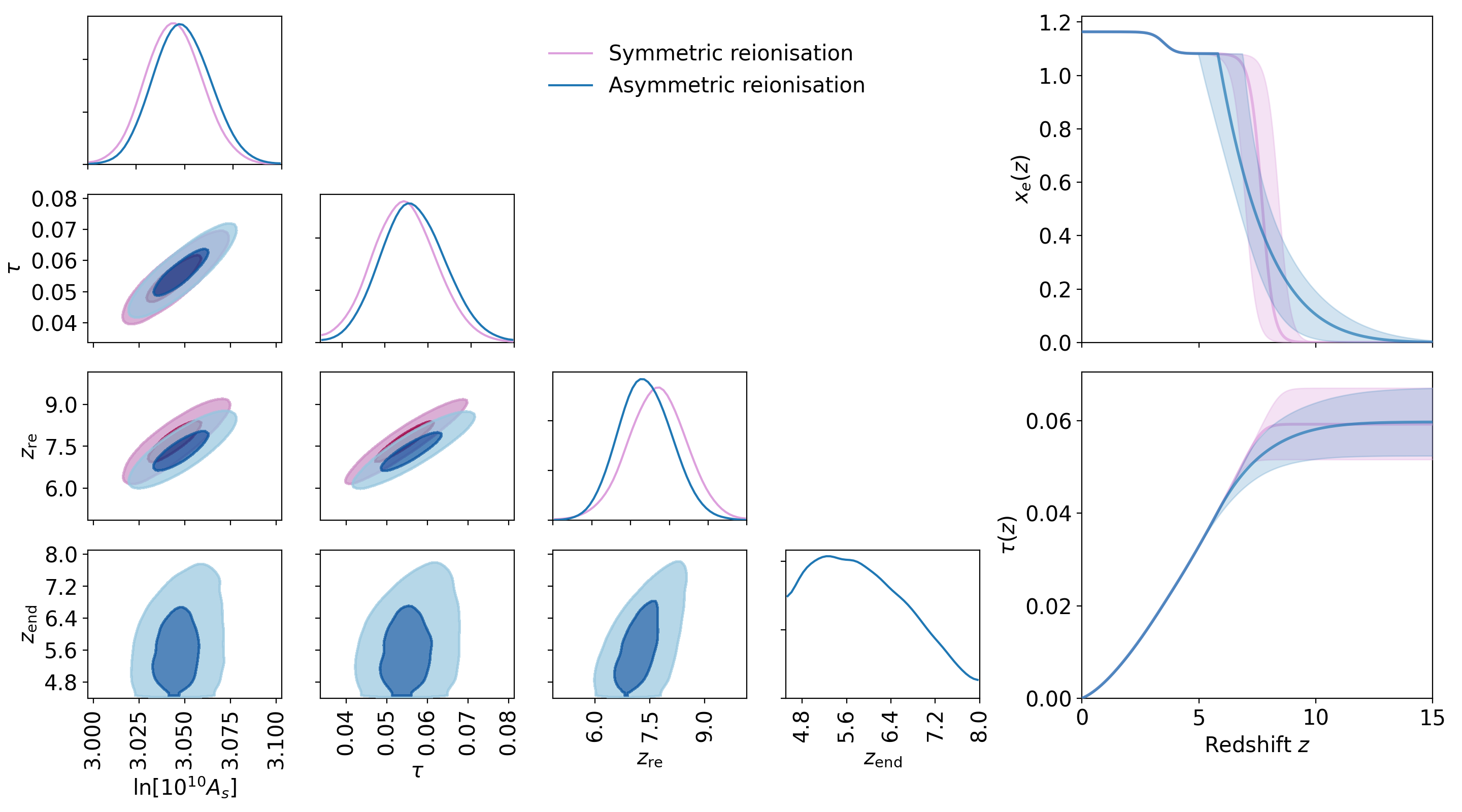}
    \caption{Results obtained when fitting the large-scale \Planck~data with two different reionisation models: a physically motivated asymmetric parameterisation of $x_e(z)$ (blue contours, Eq.~\ref{eq:xe_param}), or an instantaneous reionisation history (pink contours, Eq.~\ref{eq:xe_tanh}). \textit{Left panel:} Posterior distributions of cosmological parameters. \textit{Right panels:} Resulting redshift-evolution of the IGM ionised fraction (top) and optical depth (bottom), with associated $68\%$ confidence intervals.}
    \label{fig:res_large_scale}
\end{figure*}

\begin{table}
\caption{Maximum likelihood parameters obtained with \Planck~CMB data when an asymmetric parameterisation of reionisation is used (see text). The lower part of the table corresponds to derived parameters. Error bars are $68\%$ confidence intervals.} 
\label{tab:results_planck_asym}
\centering          
\begin{tabular}{l | c } 
Parameter & ML value \\ 
\hline  \hline
$\Omega_{\mathrm{b}}h^2$  & $0.0224 \pm 0.0001$ \\
$\Omega_{\mathrm{c}}h^2$ & $0.1199 \pm 0.0012$\\
$100 \theta_\mathrm{MC}$ & $1.0409 \pm0.0003$  \\
$n_s$ & $0.9655 \pm0.0040$ \\
$\ln(10^{10}A_s)$ & $3.048\pm 0.014$  \\
$\zre$ & $7.4 \pm 0.7$ \\
d$z$ & $1.4^{+0.7}_{-0.9}$  \\
\hline
$\tau$ & $0.056 \pm 0.007$ \\
$\zend$ & $5.8^{+0.6}_{-1.2}$  \\
\end{tabular}
\end{table}

First, we study the impact of changing the way reionisation is modelled when analysing the \citet{Planck2018} data described in Sec.~\ref{subsec:data_planck}. We compare the results obtained using the common instantaneous parameterisation of the reionisation history given in Eq.~\eqref{eq:xe_tanh}, on one side, and using the physically motivated asymmetric parameterisation given in Eq.~\eqref{eq:xe_param}, on the other side. We sample the following cosmological parameters: $\log A$, $\Omega_{\mathrm{b}}h^2$, $\Omega_{\mathrm{c}}h^2$, $100~\theta_\mathrm{MC}$ and $n_s$. 
Additionally, we sample the parameters related to the history of reionisation, given in Eq.~\eqref{eq:xe_param}, that is $\zre$ and $\zend$, with a prior $\zend > 4.5$. Here, $\tau$ is a derived parameter since the reionisation history is directly probed. Templates are used for all the foregrounds. 

The recovered maximum likelihood parameter values, along with their $68\%$ confidence intervals, are given in Table~\ref{tab:results_planck_asym}.
We find that changing the reionisation model has no impact on the baryon and dark matter densities, as well as on $100~\theta_\mathrm{MC}$. As observed in previous works \citep[e.g.][]{Planck_2016_reio,Planck2018_cosmo}, we exactly recover the cosmological constraints obtained in \citet{Planck2018} with the \texttt{tanh} model, that is $\Omega_b h^2=0.0224 \pm 0.0001$, $\Omega_c h^2 = 0.1199 \pm  0.0012$ and $100~\theta_\mathrm{MC} = 1.0409 \pm 0.0003$ at 1$\sigma$. Figure~\ref{fig:res_large_scale} compares the posterior distributions of the other four cosmological parameters sampled, for the two analyses. We see that the correlation between $\tau$ and $A_s$, commonly observed in CMB analyses, is here translated into a correlation with $\zre$. Overall, results are similar for the two cases, since the low-$\ell$ CMB data are only really sensitive to the value of $\tau$ and not to the details of the reionisation history, a point we further discuss in Sec.~\ref{sec:discussion}. Therefore, the errors on the optical depth are unchanged, as already observed by \citet{Planck_2016_reio} with \Planck~2015 data. We present the most likely reionisation history, as well as the redshift-evolution $\tau(z)$, for each case, in the right panels of Fig.~\ref{fig:res_large_scale}. In the asymmetric case (Eq.~\ref{eq:xe_param}), the extended tail towards high redshifts is compensated by a later reionisation, and a lower $\zre$, in order to reach an integrated $\tau$ value similar to the one obtained with the \texttt{tanh} model. This results in a completely different picture of the early Universe, with the first light sources starting reionising the IGM as early as $z=13$. The first half of the reionisation history is well constrained, with $\zre = 7.4 \pm 0.7$, whilst there is a lot of uncertainty on the end of the process: $\zend =  5.8^{+0.6}_{-1.2}$. However, we can place an upper limit of $\zend<7.5$ at the $95\%$ confidence level.
These results show that a clear picture of reionisation cannot be obtained with large-scale CMB data only, but must be complemented by the cosmological information enclosed in small-scale data and foregrounds.

\subsection{Substituting the kSZ template in high-$\ell$ data}
\label{subsec:res_small_scale}
\subsubsection{Fixed cosmology}

We now consider measurements of the CMB temperature fluctuations power spectrum on small scales obtained by the SPT and change the way the kSZ and tSZ signals are modelled in the analysis. Their angular power spectra are estimated using random forests (RF) trained on $C_\ell$ values computed numerically from a set of cosmological, cluster physics and reionisation parameters (see \citetalias{PAPER1}, Sec.~\ref{subsec:1_kSZ} and Appendix~\ref{app:ML}): The kSZ shape parameters, $\kappa$ and $\log\alpha_0$, as well as the mass bias $1-b$, are now let free to vary on a range where the random forests have been trained. All the sampled parameters, including $\zre$ and $\mathrm{d}z$, are listed in Table~\ref{tab:results}. In order to compare with the results of \citet{Reichardt2020}, we fix the cosmology to the values used by the authors: $\Omega_c h^2 = 0.1193 $, $\Omega_b h^2 = 0.0224$, $n_s = 0.9665$, $\log A = 3.047 $ and $100 \theta_\mathrm{MC} = 1.04101$. Here, it is not possible to fix the value of the Thomson optical depth as done in \citetalias{Reichardt2020} since, as described in the previous paragraph, the reionisation model has been altered to make $\tau$ a derived parameter. Instead, we impose a narrow Gaussian prior $\tau = 0.056 \pm 0.007$, the width of which corresponds to the errors on $\tau$ coming from large-scale data (see Sec.~\ref{subsec:res_planck_asym}). \\

\begin{figure}
    \centering
    \includegraphics[width=\columnwidth]{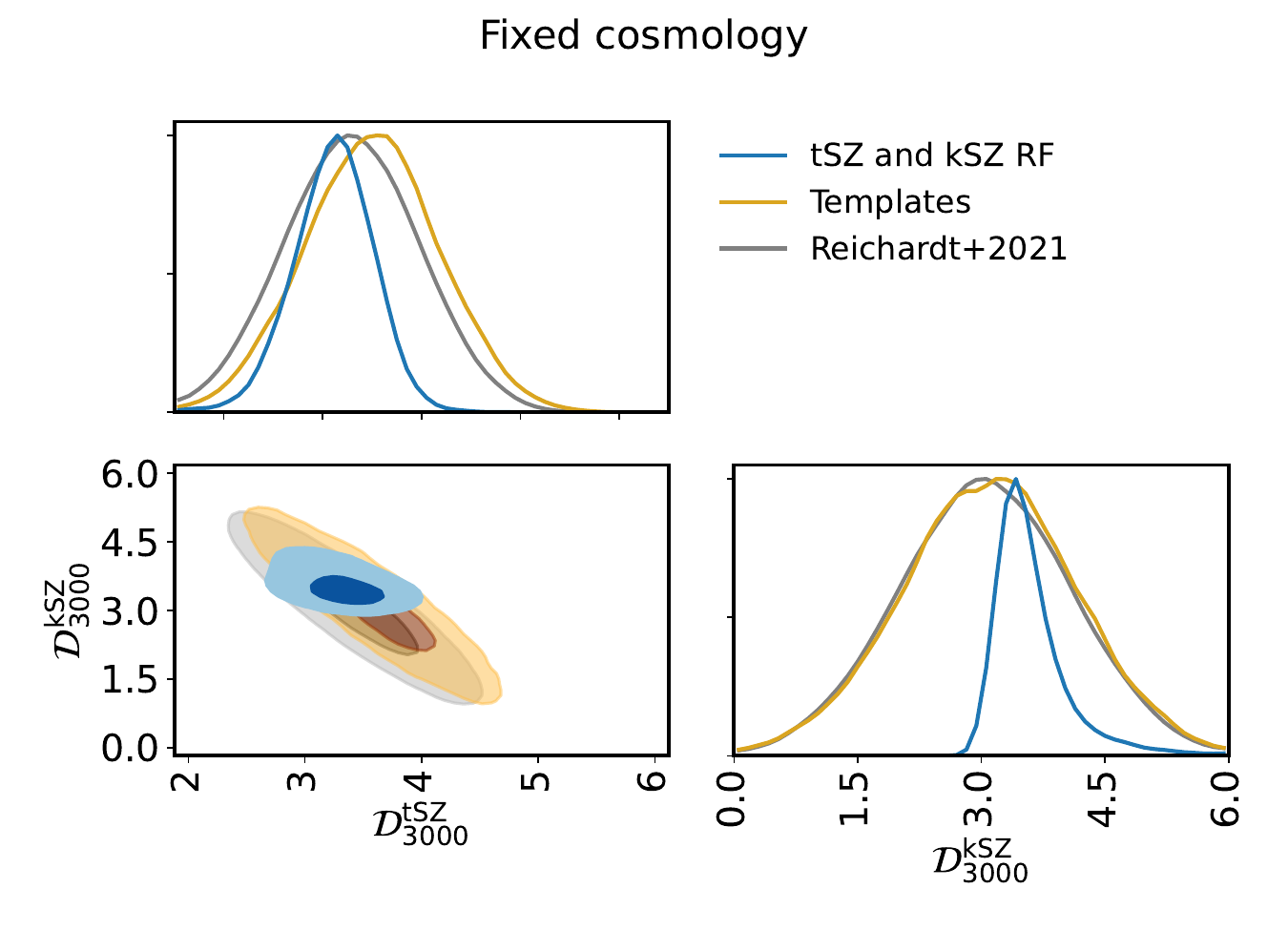}
    \includegraphics[width=\columnwidth]{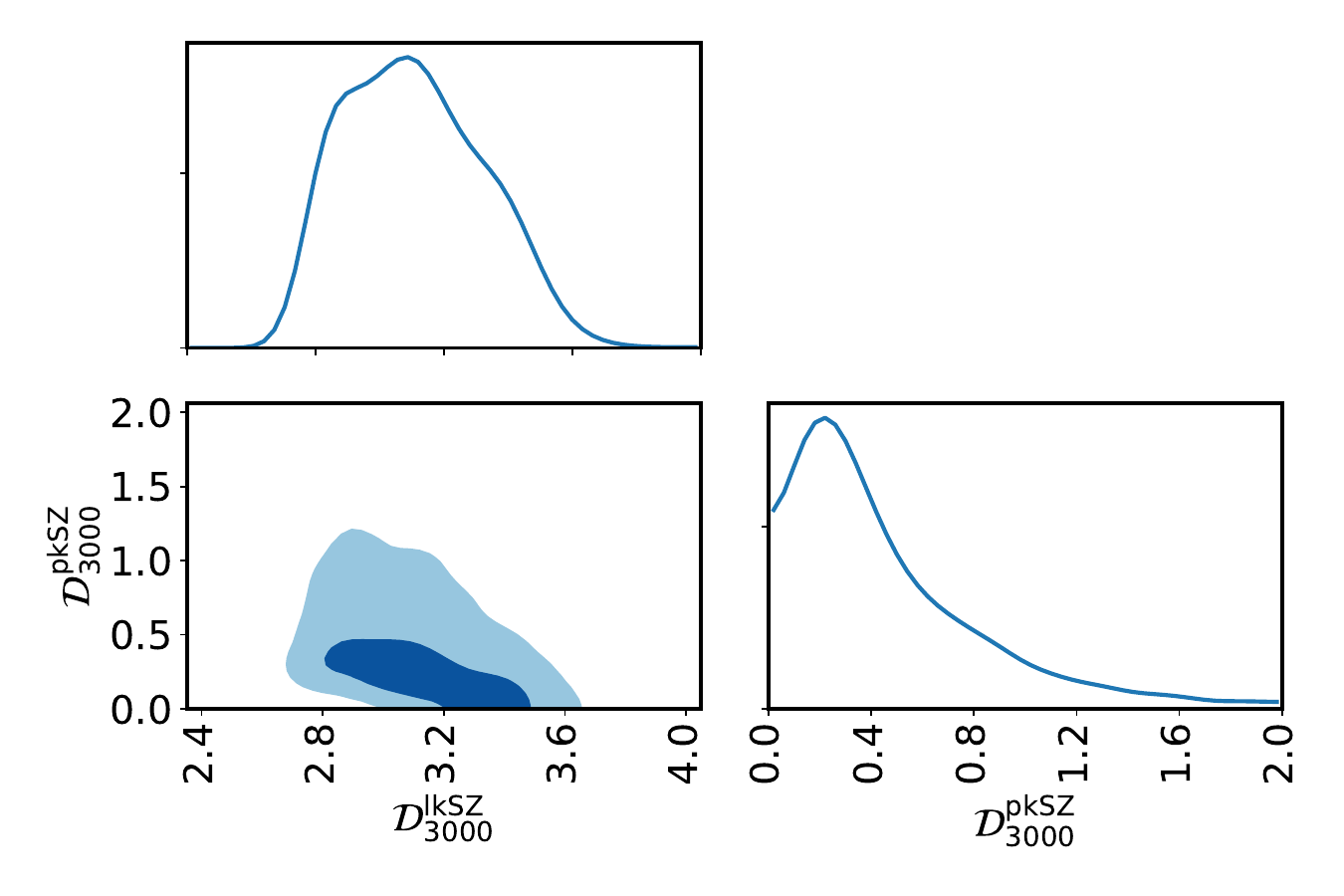}
    \caption{Constraints on the amplitude of the tSZ and kSZ power spectra at $\ell=3000$ obtained with SPT high-$\ell$ data \citepalias{Reichardt2020} using either templates (in yellow) or RF-inferred tSZ and kSZ power spectra (in blue), as well as a physical parameterisation of reionisation. Inferences are compared to the initial analysis, with templates and fixed reionisation (in grey). For all cases, the cosmology is fixed (see text for values). On the bottom panel, the kSZ is decomposed into its late-time and patchy components.}
    \label{fig:atsz_aksz_templates}
\end{figure}

We compare three cases. First, we replicate the results of \citetalias{Reichardt2020}, fixing cosmology and, effectively, reionisation. We then change the reionisation model for the one in Eq.~\eqref{eq:xe_param} but keep templates for the tSZ and kSZ spectra and sample their amplitudes at $\ell=3000$. Finally, we replace the templates by the RF-inferred tSZ and kSZ spectra, predicted at each iteration of the sampler. 

\begin{table}
    \caption{Most likely values of the amplitude, in $\mu\mathrm{K}^2$, of the SZ effects obtained when fitting the SPT data. Different cases are compared: the initial analysis by \citet{Reichardt2020}, augmented by the asymmetric reionisation model (`Temp.'), both using templates for the tSZ and the kSZ. `RF' corresponds to the use of RF-inferred, cosmology-dependent tSZ and kSZ spectra. In all cases, the cosmology is fixed.}
    \label{tab:amp_fixed_cosmo}
    \centering
    \begin{tabular}{l||c|c|c}
         & R2021 & Temp. & RF \\
         \hline
        $\mathcal{D}_{3000}^\mathrm{tSZ}$ & $3.4\pm 0.5$ & $3.6 \pm 0.5$ & $3.3\pm0.3$ \\
        $\mathcal{D}_{3000}^\mathrm{kSZ}$ & $3.1\pm 1.0$ & $3.1^{+1.0}_{-1.1}$ & $3.5^{+0.4}_{-0.2}$ \\
        \hspace{2mm} patchy & -- & -- & $<1.5$ \\
        \hspace{2mm} late-time & -- & -- & $ 3.1^{+0.1}_{-0.3}$
    \end{tabular}
\end{table}

In the top panel of Fig.~\ref{fig:atsz_aksz_templates}, we show the resulting posterior distributions of the amplitudes of the tSZ and total kSZ power spectra at $\ell=3000$ and $143~\mathrm{GHz}$, labelled as $\mathcal{D}_{3000}^\mathrm{tSZ}$ and $\mathcal{D}_{3000}^\mathrm{kSZ}$, respectively. Maximum likelihood values, along with $68\%$ credibility intervals, are given in Table~\ref{tab:amp_fixed_cosmo}.
Similarly to what was observed in \citetalias{PAPER1}, including the cosmology dependence of the SZ spectra in the analysis limits the values their amplitudes can take and breaks their degeneracy, leading to tighter constraints. We move from a $3\sigma$ measurement reported in \citetalias{Reichardt2020} to a $6\sigma$ measurement of the amplitude of the kSZ signal and the error bars are cut in half: $\mathcal{D}_{3000}^\mathrm{kSZ} = 3.5^{+0.4}_{-0.2}~\mu\mathrm{K}^2$. The RF prediction of the kSZ signal allows to decompose it between the contribution of post-reionisation (late) and patchy signals. In the bottom panel of Fig.~\ref{fig:atsz_aksz_templates}, we present the probability distributions of their amplitude at $\ell=3000$. The former largely dominates the total kSZ power, contributing to about $86\%$ at $\ell=3000$ with $\mathcal{D}_{3000}^\mathrm{lkSZ}=3.1^{+0.1}_{-0.3}~\mu\mathrm{K}^2$. Because of its low amplitude, only an upper limit can be placed on the latter in this setting: $\mathcal{D}_{3000}^\mathrm{pkSZ}<1.5~\mu\mathrm{K}^2$ at 95\% confidence.
Regarding the tSZ amplitude, the uncertainty is reduced by $40\%$ compared to \citetalias{Reichardt2020}, and we move from a $6$ to a $10\sigma$ measurement, such that $\mathcal{D}_{3000}^\mathrm{tSZ} = 3.3 \pm 0.3~\mu\mathrm{K}^2$. The distributions of other foreground
parameters, shown in Fig.~\ref{fig:app_nuisance_reichardt_comparison}, are also impacted by the change in the tSZ and kSZ modelling. For example, the amplitude of the contribution of dusty galaxies to the observed temperature power spectrum is better constrained, whilst staying consistent with previous results. The constraints on the amplitude of the tSZ-CIB correlations do not change whether we use templates or not ($\xi = 0.07 \pm 0.03$), although the degeneracy of this parameter with $\mathcal{D}_{3000}^\mathrm{kSZ}$ and $\mathcal{D}_{3000}^\mathrm{tSZ}$ is lifted when templates are replaced by RF predictions. In agreement with \citetalias{Reichardt2020}, the data favour positive values of $\xi$, and a negative tSZ-CIB correlation, corresponding to an overdensity of dusty galaxies in galaxy clusters.

The cosmological information included in the high-$\ell$ SPT data is sufficient to give good constraints on the midpoint of reionisation and we find that, for a fixed cosmology, the use of a full derivation of the SZ power spectra instead of templates does not help constraining any of the reionisation parameters better. That is because changing the model but fixing the cosmology is equivalent to using another template, which \citetalias{Reichardt2020} has shown to have no impact on the results. The model favoured by the small-scale data in this setting is one of early reionisation ($\zre  = 8.5^{+0.5}_{-0.4}$), in tension with currently available data points on the ionisation fraction of the IGM (see Fig.~\ref{fig:xe_spt}) and can be traced back to large optical depth values ($\tau = 0.069\pm 0.004$), over $3\sigma$ away from the centre of the Gaussian prior imposed to the fit, and from the \citet{Planck2018} constraints. This is due to the fact that on the $\ell$-range covered by the SPT data, there is a strong correlation between the $A_s$ and $\tau$ parameters, as the amplitude of the primary temperature power spectrum follows $A_s ~\mathrm{exp}(-2\tau)$. Fixing $A_s$ is therefore equivalent to fixing $\tau$ and the chosen value of $A_s$ corresponds to a large value of $\tau$, which in turn triggers early reionisation scenarios and large kSZ amplitudes. We confirm this intuition by running the MCMC with RF-predicted tSZ and kSZ spectra and fixed cosmology, only changing the value of $A_s$ to a lower value. This set-up leads to lower values of $\tau$, and, in turn, of the reionisation midpoint and of the patchy kSZ amplitude.

\subsubsection{Free cosmology}

From the results presented in previous section, it is clear that the constraints on SZ amplitudes and on reionisation obtained with the SPT data strongly depend on the assumed cosmological model. Therefore, we now let cosmological parameters free but apply tight Gaussian priors corresponding to the posterior distributions of \citet{Planck2018} on $\Omega_b h^2$, $\Omega_c h^2$ and $\theta_\mathrm{MC}$, which we have found in Sec.~\ref{subsec:res_planck_asym} to be largely impervious to changes in the reionisation history. On the other hand, $\log A$ is let completely free, along with the reionisation parameters and the mass bias $1-b$, related to the computation of the tSZ spectrum\footnote{As shown in \citetalias{PAPER1}, it is useful to keep in mind that the bias $1-b$ is strongly correlated to the amplitude of the tSZ power spectrum at $\ell=3000$ \citep[see also][]{Salvati2019}.}.
The applied priors are listed in Table \ref{tab:results}.
To assess the constraining power of our newly introduced cosmology-dependent tSZ and kSZ power spectra, we compare several cases, where either: Templates are used for both SZ spectra, a template is used for one SZ effect, whilst the other is predicted by the RF, or both SZ spectra are predicted by the RF. 
The latter set-up leads to the main results of this paper.
The full list of maximum likelihood parameters resulting from these four analyses is given in Table~\ref{tab:results} and a complete triangle plot of the posterior distributions of all parameters sampled, apart from nuisance parameters, can be found in Fig.~\ref{fig:full_triangle_plot}.

\begin{figure}
    \centering
    \includegraphics[width=\columnwidth]{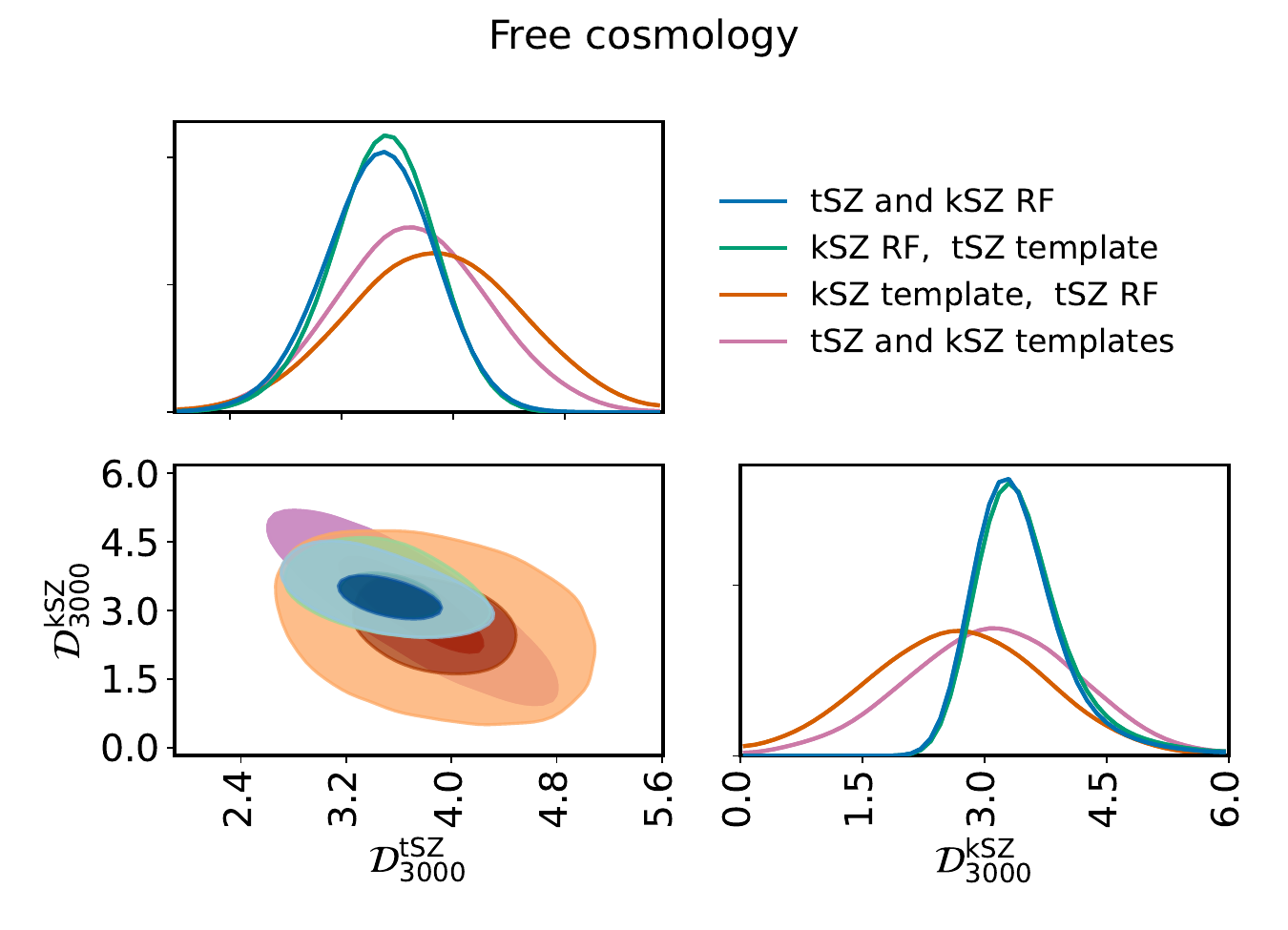}
    \includegraphics[width=\columnwidth]{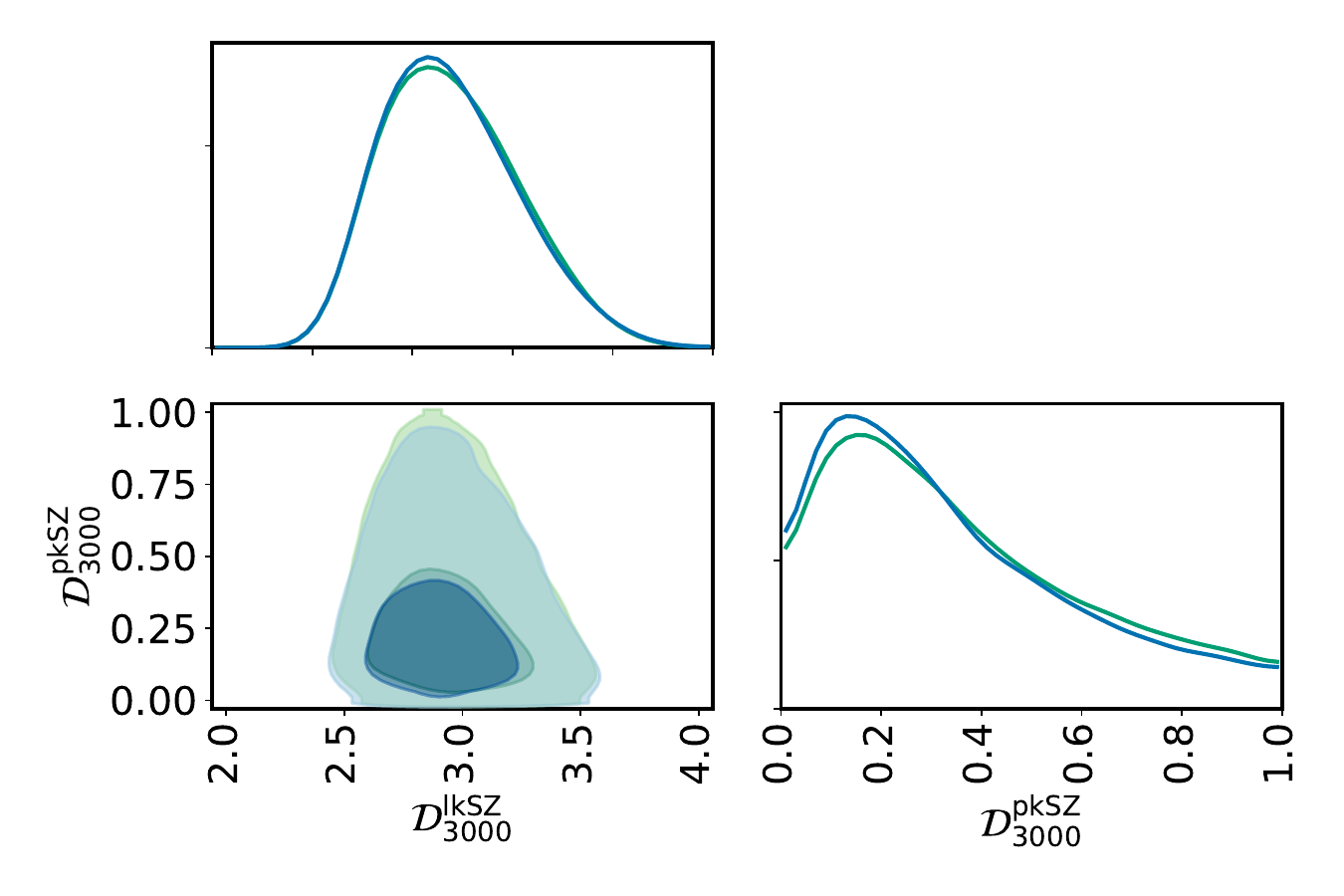}
    \caption{Probability distributions of the amplitudes of the tSZ and kSZ power spectra at $\ell=3000$ for the different models considered (see text). On the bottom panel, the kSZ is decomposed into its late-time and patchy components.}
    \label{fig:triangle_atksz_free_cosmo}
\end{figure}
\begin{figure}
    \includegraphics[width=.95\columnwidth]{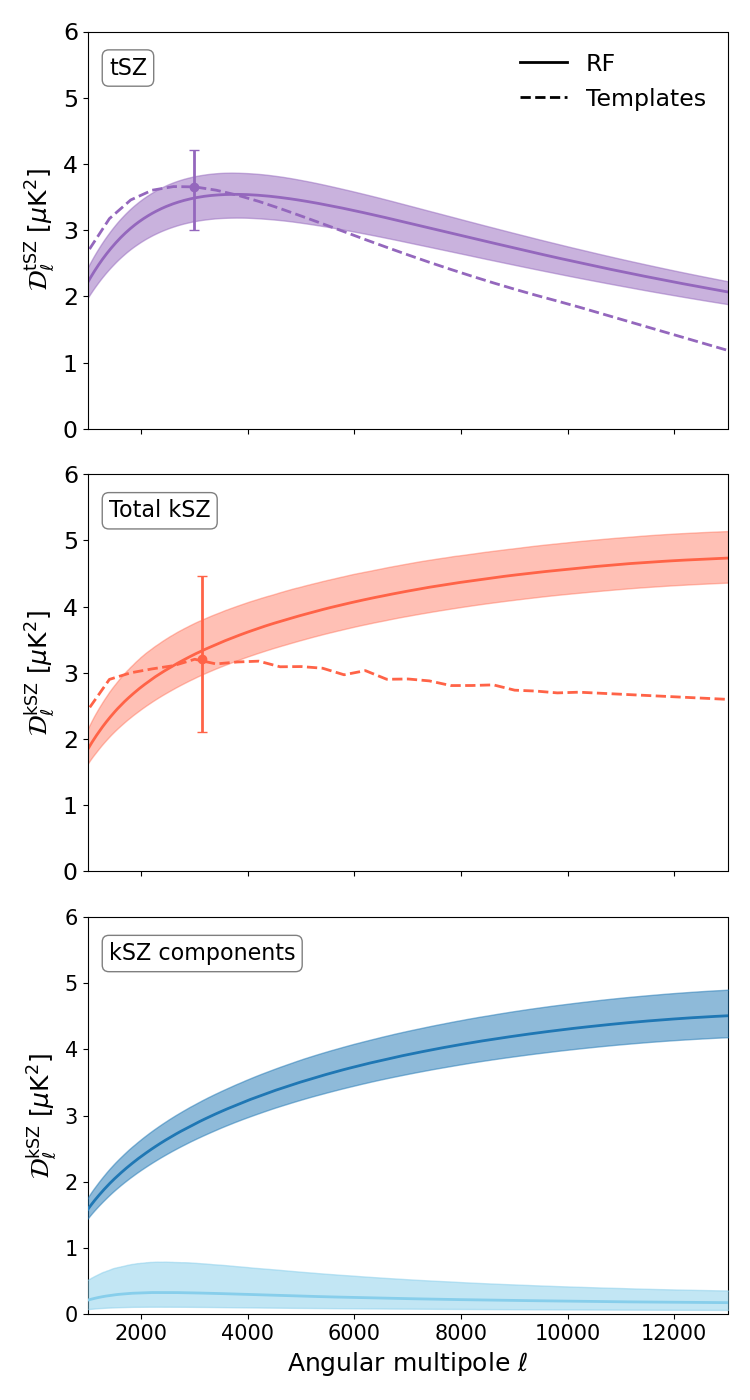}
    \caption{Most likely SZ spectra at $150~\mathrm{GHz}$ obtained when fitting cosmology and reionisation parameters to the SPT data. Two cases are compared: Either templates are used for both spectra (dashed lines) or they are predicted by our RF given the sampled parameters (solid lines). For the results obtained with RF predictions, shaded areas correspond to $68\%$ confidence regions. For the results obtained with templates, only the amplitudes at $\ell=3000$ are constrained, and the $68\%$ confidence intervals on their values are shown as error bars. The tSZ and total kSZ spectra are shown in the upper and middle panels, respectively. When possible, the kSZ signal is decomposed between patchy (dark blue) and late-time (light blue) components, shown in the lower panel.}
    \label{fig:sz_ps}
\end{figure}
\begin{figure*}
    \centering
    \includegraphics[width=.9\textwidth]{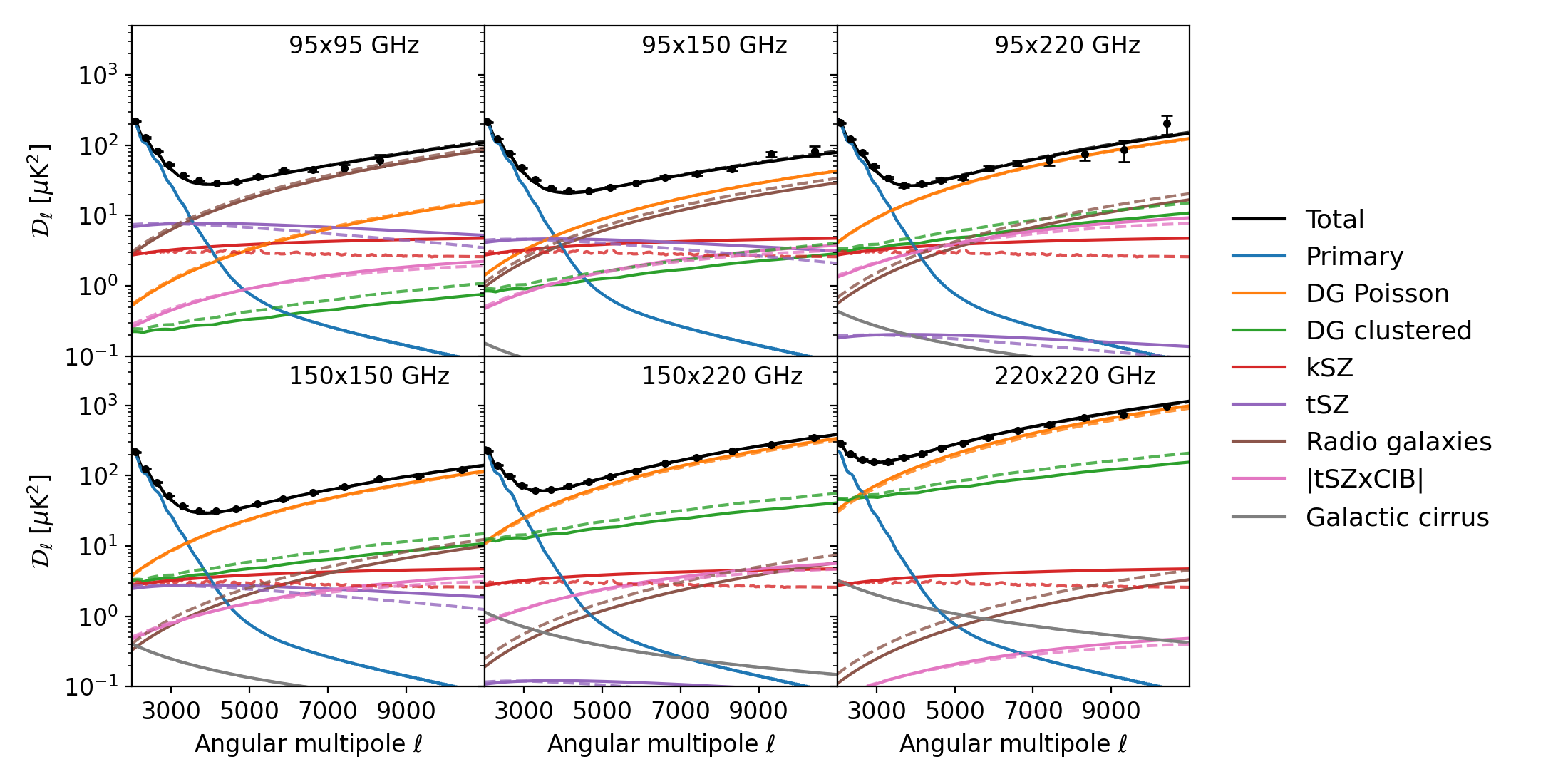}
    \caption{Most likely components of the small-scale CMB temperature power spectrum observed by the SPT \citepalias{Reichardt2020}, when cosmology is free. Solid and dashed lines show the results when RF predictions and templates are used for the SZ spectra, respectively.}
    \label{fig:spt_components}
\end{figure*}

In the upper panel of Fig.~\ref{fig:triangle_atksz_free_cosmo}, we show the posterior probability distributions of the tSZ and total kSZ amplitudes at $\ell=3000$ for the four considered cases. Again, replacing one of the two SZ templates by a cosmology-dependent spectrum is sufficient to break the degeneracy between the two amplitudes, as they can only take values compatible with the cosmology. However, here, replacing the tSZ template but keeping the kSZ one does not improve detection limits, with $\mathcal{D}^{\rm kSZ}_{3000}$ remaining at $2\sigma$. This can be related to the strong priors placed on cosmological parameters, in particular on $\Omega_c h^2$ and, therefore, on $\Omega_m$, almost fixing the shape of the tSZ spectrum. Because of the strong correlation existing between $\Omega_c h^2$ and the mass bias $1-b$, releasing the prior on $\Omega_c h^2$ will weaken the constraints obtained on the mass bias, but have no impact on the measured $\mathcal{D}_{3000}^\mathrm{tSZ}$. This also explains why the results only slightly change when the tSZ model changes but RF-inferred kSZ is used. 

It is only when the cosmological information included in the kSZ spectrum is used that there is a clear improvement in the measurement of the two signals, with the most likely values of the tSZ and kSZ amplitudes sitting at $9$ and $5\sigma$, respectively, from a zero amplitude: $\mathcal{D}_{3000}^\mathrm{tSZ} = 3.5 \pm 0.4$ and $\mathcal{D}_{3000}^\mathrm{kSZ}=3.4^{+0.5}_{-0.3}$ with a $68\%$ confidence limit (see also Table~\ref{tab:results}). The model used to derive the kSZ allows to decompose it into its patchy and late-time contributions. The probability distributions of their amplitudes at $\ell=3000$ are shown in the bottom panel of Fig.~\ref{fig:triangle_atksz_free_cosmo}. Again, if the late-time contribution is well-constrained, only an upper limit can be placed on the patchy kSZ, similar to what was obtained with a fixed cosmology: $\mathcal{D}_{3000}^\mathrm{pkSZ} < 1.6~\mu\mathrm{K}^2$ at 95\% confidence. The most likely kSZ and tSZ spectra at $150~\mathrm{GHz}$, as well as their $68\%$ credibility regions, are shown in Fig.~\ref{fig:sz_ps} and compared to the best-fit templates. For the first time, it is possible to see the shape of the SZ spectra preferred by the small-scale CMB data, and to decompose the kSZ power into its late-time and patchy components. In conclusion, in comparison to the templates, the RF-prediction allows for a larger power at high-$\ell$ for both spectra and for reduced error bars at all multipoles. This result holds when using different templates than the ones used in \citet{Reichardt2020}, notably templates obtained with the analytical derivation of \citet{Gorce2020} for different sets of parameters. We additionally find a difference of only $0.17\,\mu\mathrm{K}^2$, corresponding to approximately $0.25\sigma$, on the amplitude of the total kSZ power at $\ell=3000$ when a template and a cosmology-dependent amplitude are used for the late-time component of the kSZ \citep{ShawRudd_2012}, whilst the RF is used to infer the patchy contribution.

We show in Fig.~\ref{fig:spt_components} the maximum likelihood components of the temperature power spectrum observed by the SPT \citepalias{Reichardt2020} for our main analysis, that is when both SZ spectra are predicted by the RF, and when templates are used. We see that the SZ amplitudes at $\ell=3000$ are consistent across all analyses, but that freeing the shape of the SZ spectra when introducing the RF leads to an increase in the contribution of both the kSZ and the tSZ to the total power on the smallest scales ($\ell>4000$). This power is mostly drawn from clustered dusty galaxies and radio galaxies. The fixed shape of their spectra results in decreased amplitudes at $\ell=3000$, with the amplitude of the contribution from radio galaxies moving from $\mathcal{D}_{3000}^\mathrm{rg}=0.9\pm 0.2$ to $\mathcal{D}_{3000}^\mathrm{rg}=0.7\pm0.1$ when introducing the RF. The amplitudes of the other foregrounds are only weakly impacted by the new model. For example, the amplitude of the contribution from the Galactic cirrus at $\ell=3000$ and $220~\mathrm{GHz}$ is $\mathcal{D}_{3000}^\mathrm{gc} = 1.9\pm 0.4~\mu\mathrm{K}^2$.  Only the constraints on the CIB-tSZ correlation, whose amplitude is described by a sampled parameter $\xi$ (see Sec.~\ref{sec:data}), are strongly impacted by the change of model. We show in Fig.~\ref{fig:tSZxCIB_free_cosmo} the joint posterior distributions of this parameter with the kSZ and tSZ amplitudes at $\ell=3000$ in the four cases aforementioned. In the analysis of previous data, allowing for a correlation between the two signals has made measuring the kSZ signal more difficult \citep{reichardt_2012_spt,DunkleyCalabrese_2013,SieversHlozek_2013,GeorgeReichardt_2015}, until \citetalias{Reichardt2020}, where the authors claim to measure the kSZ amplitude at $3\sigma$, despite letting $\xi$ free. The previous difficulties to measure the SZ amplitudes are explained by a strong degeneracy between $\xi$ and the kSZ and tSZ power. Here, when using both templates, we indeed find a correlation coefficient of $0.65$ and $-0.58$ between $\xi$ and $\mathcal{D}_{3000}^\mathrm{kSZ}$ and $\xi$ and $\mathcal{D}_{3000}^\mathrm{tSZ}$, respectively. However, when removing the use of templates, these coefficients are reduced to $0.22$ and $-0.22$, respectively, such that the level of correlation between the CIB and the tSZ is constrained to $\xi = 0.07 \pm 0.03$, ruling out positive correlations at the $2\sigma$ level and confirming that there is an overdensity of dusty galaxies in galaxy clusters.

\begin{figure*}
    \centering
    \includegraphics[width=\textwidth]{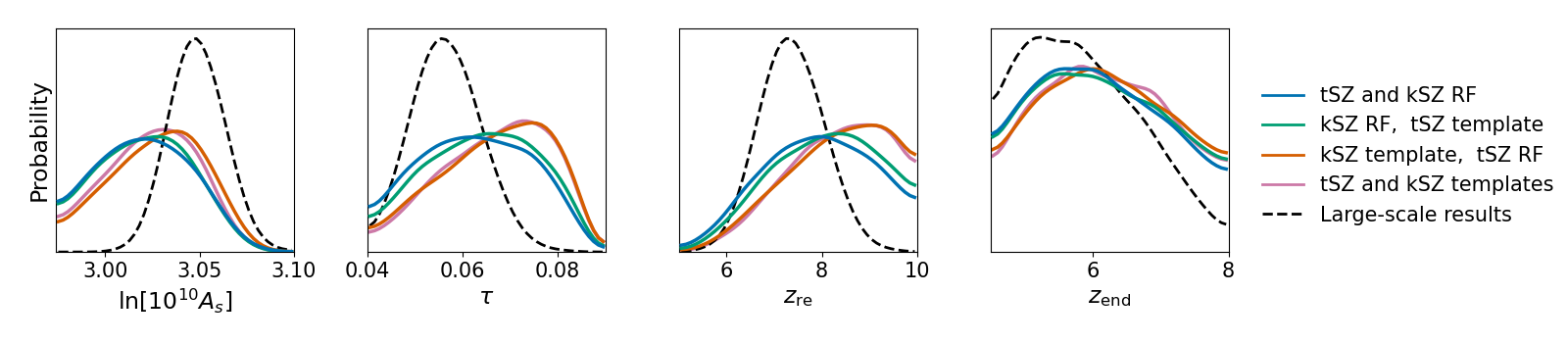}
    \caption{Posterior distributions of cosmological and reionisation parameters sampled when cosmology is let free with \citet{Planck2018} priors and RF-predicted or templates are used for the kSZ and tSZ power spectra. The constraint obtained with large-scale \Planck~data in Sec.~\ref{subsec:res_planck_asym} are shown as black dashed lines.}
    \label{fig:triangle_reion_free_cosmo}
\end{figure*}

Regarding cosmological constraints, as shown in Table~\ref{tab:results}, we again recover the \Planck~priors on $\Omega_b h^2$, $\Omega_c h^2$, $100\theta_\mathrm{MC}$ and $n_s$. Overall, the values of the other sampled cosmological parameters are consistent with the large-scale results, only with larger error bars. Notably, the width of the $68\%$ confidence interval on $\log A$ increases from $0.014$ to $0.027$ when small-scale data are used in the analysis. In Fig.~\ref{fig:triangle_reion_free_cosmo}, we show the posterior distributions of $A_s$, $\tau$, $\zre$ and $\zend$ for our four cases. As before, we see that replacing the tSZ template by a cosmology-dependent spectrum does not impact the parameter constraints, because of  the strong priors placed on $\Omega_c h^2$ and so, effectively, on the shape of the spectrum. Introducing  the RF-inferred kSZ spectrum, however, slightly shifts the most likely values of $A_s$ and $\zre$, and, consequently, of $\tau$, pulling the constraints on the reionisation history towards later scenarios: $\zre$ moves from $8.4^{+0.8}_{-1.3}$ with both templates to $\zre = 7.9^{+1.1}_{-1.3}$ with both cosmology-dependent spectra, whilst $\tau$ moves from $0.068^{+0.009}_{-0.014}$ to $\tau = 0.062^{+0.012}_{-0.015}$ at the $68\%$ confidence level.

\section{Discussion}\label{sec:discussion}

The constraints on the history of reionisation derived from CMB analyses are traditionally based on the value of the Thomson optical depth which only offers a limited amount of information, since it is the integrated electron column density along the line of sight, averaged over the whole sky.
In particular, as seen in Sec.~\ref{subsec:res_planck_asym}, constraining models with more than one parameter with current CMB data proves difficult. Additionally, the measurements of $\tau$ from \Planck~data are still plagued with systematics and different analyses lead to different values \citep[e.g.][]{Planck_2016_reio, PaganoDelouis_2020, NatalePagano_2020} which, whilst being mostly consistent, can change our picture of the high-redshift Universe. In order to improve the constraints on reionisation obtained from CMB data, one can combine CMB observations to other data sets, such as measurements of the IGM ionised fraction from quasar spectra, galaxy luminosity functions \citep[e.g.][]{GorceDouspis_2018,QinPoulin_2020}, or even measurements of the 21\,cm signal \citep[e.g.][]{Alvarez_2016,AhnShapiro_2021,PaulMukherjee_2021,BillingsLaPlante_2021}. The option we chose in this work is to fully exploit the potential of CMB observations by including small-scale data in the analysis. In contrast to previous works \citep{Planck_2016_reio}, cosmological- and reionisation-dependent models of the tSZ and kSZ power spectra, which are foregrounds to the primary temperature fluctuations on scales $\ell>2000$, are introduced, giving access to the cosmological information included in their shape and amplitude. Doing so, we provide new constraints on the nature of reionisation and on cosmological parameters. 
 

\subsection{Constraining the nature of reionisation}
\label{subsec:5_xe}

In Sec.~\ref{subsec:res_planck_asym}, fitting the \Planck~data with different parameterisations of the redshift-evolution of the IGM ionised fraction (Eqs.~\ref{eq:xe_tanh} and \ref{eq:xe_param}), we have found that the large-scale CMB power spectrum can only constrain the value of the optical depth $\tau$, but not the details of the reionisation history. However, the measured value of $\tau$ can give a completely different picture of the high-redshift Universe, depending on the adopted model, and the latter should therefore be chosen carefully. A complementary analysis has been performed by \citet{QinPoulin_2020}, who directly confront semi-numerical simulations \citep{21cmFAST_2011,GreigMesinger_2015,Murray2020} to \citet{Planck2018} large-scale data, avoiding the use of any parameterisation of the reionisation history. They recover a constraint on $\tau$ similar to parametric analyses, and argue that a prior on $\tau$ can be used with the same effect as the full large-scale \Planck~likelihoods to constrain reionisation models.
Another way of obtaining somewhat model-free constraints on reionisation from CMB data is the use of non-parametric reconstructions, for example via a principal component analysis \citep{HuHolder_2003,MortonsonHu_2008,BeginLiu_2021} or by interpolating between a varying number of redshift bins along the ionisation history \citep{MilleaBouchet_2018}. 
These results motivate a more thorough analysis of small-scale data to complement large-scale results and better constrain reionisation.

\begin{figure}
    \centering
    \includegraphics[width=\columnwidth]{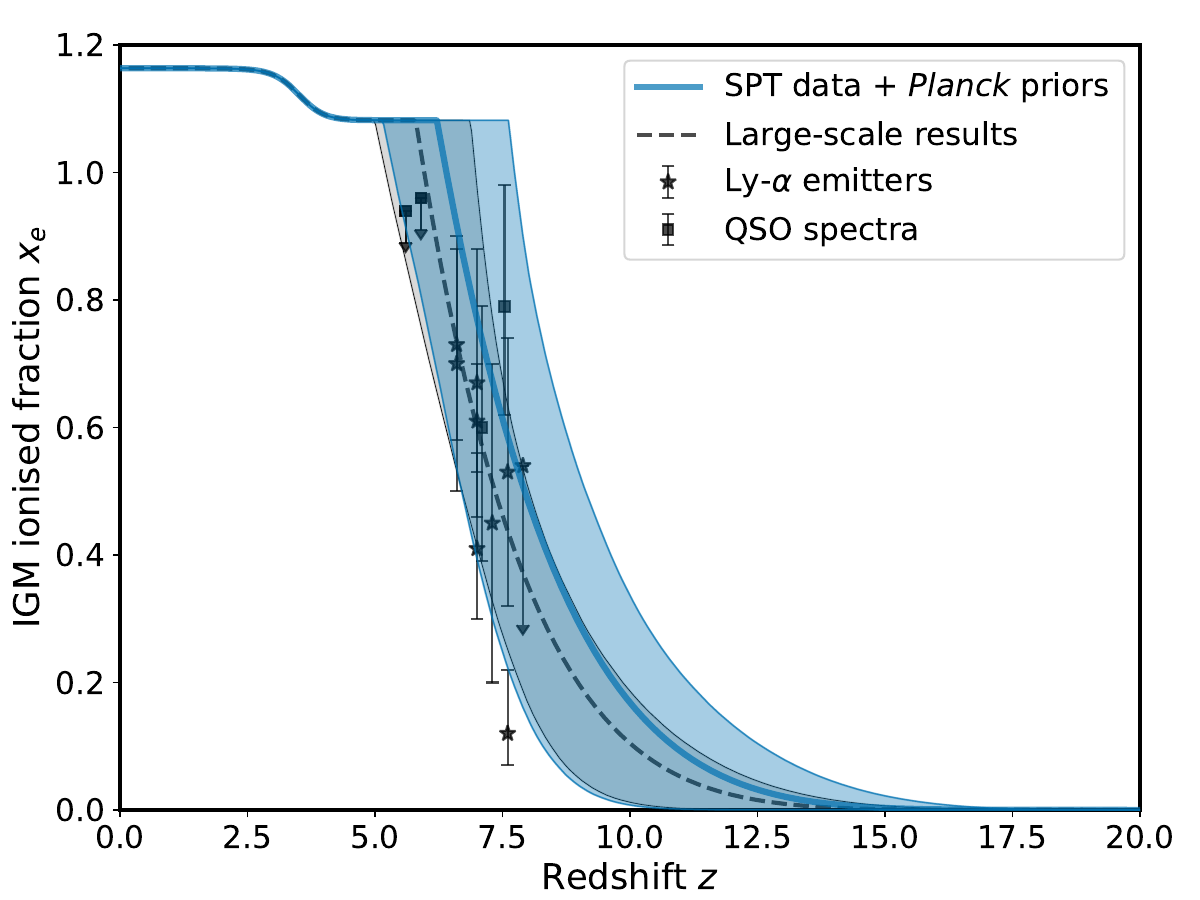}
    \caption{Constraints on reionisation history obtained with \Planck~\citep[in grey,][]{Planck2018} and SPT high-$\ell$ data \citep[in blue,][]{Reichardt2020} for cosmology-dependent kSZ and tSZ spectra. Cosmology is free but, for the latter, Gaussian priors from Sec.~\ref{subsec:res_planck_asym} are imposed on four cosmological parameters. Inferences are compared to currently available data points on the ionisation fraction of the IGM from Lyman-$\alpha$ emitters and QSO spectra (see text for references).}
    \label{fig:xe_spt}
\end{figure}

Using cosmology-dependent SZ spectra to fit the small-scale SPT data, we find in Sec.~\ref{subsec:res_small_scale} that their shape and amplitude can indeed be used to learn about the reionisation history.
In Fig.~\ref{fig:xe_spt}, we compare the results for free cosmology runs on large-scale and small-scale data, and see that both data sets tell consistent stories, in agreement with current constraints on the ionisation fraction of the IGM from Lyman-$\alpha$ emitters \citep{ota_2008,schenker_2013,konno_2017,mason_2017,HoagBradac_2019,JungFinkelstein_2020} and quasar spectra \citep{mcgreer_2015,GreigMesinger_2017,GreigMesinger_2019}, or a combination of both \citep{GorceDouspis_2018}. With large-scale data only, we report a constraint on the midpoint of reionisation of $7.4\pm0.7$. Using the cosmology-dependent kSZ and tSZ spectra to analyse the SPT data \citep{Reichardt2020}, we obtain $\zre = 7.9^{+1.1}_{-1.3}$ for the midpoint and $\zend = 6.2^{+1.6}_{-0.8}$, corresponding to a $95\%$ upper limit of $\zend < 8.5$, for the endpoint (see Fig.~\ref{fig:triangle_reion_free_cosmo}). The small-scale data requires larger values of the reionisation midpoint than large-scale results, with the first light sources starting to ionise the IGM around $z=15$\footnote{The parameterisation used allows for reionisation to start as early as $z=20$, but not before.}. This fact was already observed in \citet{ChoudhuryMukherjee_2021}, where the authors fit a semi-numerical model of reionisation to \Planck~CMB and SPT data and could mean a tension between the small-scale CMB data and recent works inferring a late end for reionisation from astrophysical data sets \citep{BeckerBolton_2015,BosmanFan_2018}. More precisely, let the duration of reionisation $\Delta_{z,50} = z_{0.25}-z_{0.75}$, where $z_{i}$ is the redshift at which a fraction $i$ of the IGM is ionised. We find an upper limit $\Delta_{z, 50} < 4.8$ ($4.3$) at $99\%$ ($95\%$) confidence, consistent with the constraints obtained in previous works from the SPT measurement of the kSZ amplitude \citep{Reichardt2020}. These include $\Delta_{z, 50} < 2.9$ ($99\%$ C.L.) presented in \citet{ChoudhuryMukherjee_2021} and $\Delta_{z, 50} < 3.5$ given in \citet{ChenTrac_2022}, although we expect the latter limit to be significantly loosened when the parameters of the model used by the authors are freed. Our upper limit is also less stringent that the initial results of \citet{Planck_2016_reio}, where the authors post-process the posterior distribution of $\DkSZ$ obtained with templates confronted to a combination of \Planck, ACT and SPT data \citep{DasLouis_2014, GeorgeReichardt_2015, PlanckCollaborationAdam_2016}. We add that, as mentioned in Appendix~\ref{app:ML}, kSZ measurements alone seem unsufficient to obtain precise constraints on $\zre$, and a coherent analysis of small- and large-scale data is required.

The new analytic derivation of the kSZ angular spectrum introduced in \citet{Gorce2020} and confronted to data for the first time in this work does not only allow a consistent modelling of the reionisation history in large and small CMB data analysis, but also gives access to information about the patchiness of reionisation and the nature of its sources. Indeed, the $\kappa$ parameter is directly related to the typical size of ionised bubbles, whilst $\alpha_0$ describes the variance of the electron density field during the EoR, and the distribution of light sources throughout the IGM.
However, current data does not allow to get precise constraints on these two shape parameters. We can report an upper limit $ \alpha_0 < 10^{4.1}~\mathrm{Mpc}^3$ at 95\% confidence, which is in good agreement with values obtained for both radiative hydrodynamics and semi-numerical simulations of reionisation \citep{21cmFAST_2011,aubert_2015_EMMA,rsage} in \citet{Gorce2020}. We expect the results presented here to greatly improve with data from the next generation of CMB observatories. Additionally, these parameters being related to the morphology of reionisation, and so to the physical properties of early galaxies leading the process, future observations, for example of the spatial fluctuations of the 21cm signal by the Square Kilometre Array \citep{MellemaKoopmans_2015}, will give access to independent constraints on these parameters, and therefore help the characterisation of the high-multipole CMB power spectrum.


\subsection{Cosmological constraints}
\label{subsec:5_cosmo}

In Fig.~\ref{fig:triangle_reion_free_cosmo}, we compare the posterior distributions of $\log A$, $\tau$, $\zre$ and $\zend$ obtained in Sec.~\ref{subsec:res_planck_asym} with \Planck~data only, and in Sec.~\ref{subsec:res_small_scale} with the SPT data, \Planck~priors and RF-inferred SZ spectra. We see that, because of the presence of many different components to the observed temperature power spectrum on small scales, the SPT data are less constraining than \Planck~CMB. In particular, the $1\sigma$ intervals on $\log A$ and $\tau$ close to double, from $0.007$ reported in Sec.~\ref{subsec:res_planck_asym} to $0.012$ for the latter. Interestingly, the $\tau-A_s$ degeneracy is not broken by the new reionisation model and the addition of the SPT data. However, we expect that future experiments will be able to break this degeneracy \citep{CalabreseHlozek_2014,GalliBenabed_2014} and such forecasts will be the focus of future work.
The degeneracy between the two parameters, although corresponding to the same correlation coefficient, is shifted between the large- and small-scale results. That is, for a given $\log A$, large-scale data favours larger $\tau$ values than small-scale data, and reciprocally. This is due to the small-scale data requiring a kSZ amplitude corresponding to early reionisation scenarios, which in turn translates into a larger optical depth. Hence, the best-fit value for $\log A$ is found to be smaller than, although still fully consistent with, the large-scale results. 

Another potential issue of this analysis is the strong correlation observed between the reionisation midpoint $\zre$ and the Thomson optical depth $\tau$, which is directly derived from the reionisation history. Indeed, we report a correlation coefficient between the two parameters of $0.98$. This degeneracy could be partly broken with future data to allow for a cleaner measurement of both reionisation history parameters, $\zre$ and $\zend$. For example, measurements of the spatial fluctuations of the CMB optical depth could be used \citep{HolderIliev_2007,DvorkinSmith_2009,GluscevicKamionkowski_2013,NamikawaRoy_2021}. Another indirect option would be to independently constrain $\zre$ and $\zend$ with data from other cosmological sources, such as quasar spectra or the power spectrum of fluctuations in the 21\,cm signal coming from the EoR \citep{LiuPritchard_2016} or with next generation CMB data. Indeed, \citet{AlvarezFerraro_2021} suggest that varying the reionisation history has a different impact on the Gaussian and the non-Gaussian part of the signal\footnote{For example, the kSZ four-point correlations will be enhanced by shorter reionisation histories.}, and, therefore, four-point estimators of the kSZ could be used to break the degeneracy.\\

In order to break degeneracies between parameters, detect systematics, and allow a fruitful analysis of available data sets, a consistent analysis of the different observables of reionisation seems to be a promising avenue. Indeed, several works have already demonstrated the potential offered by a combined analysis of astrophysical and CMB data \citep[e.g.][]{MilleaBouchet_2018,QinPoulin_2020,ChatterjeeChoudhury_2021,BaxterWeinberger_2021,Sato-PolitoBernal_2021,LaPlanteSipple_2022}, 21\,cm and CMB data \citep[e.g.][]{TashiroAghanim_2011,Alvarez_2016,MaHelgason_2018,HotinliJohnson_2020,LaPlanteLidz_2020,BillingsLaPlante_2021,ChoudhuryMukherjee_2021,AhnShapiro_2021,BeginLiu_2021}, or different types of CMB data \citep[e.g.][]{Planck_2016_reio,AlvarezFerraro_2021,NamikawaRoy_2021,PaulMukherjee_2021,RoyKulkarni_2021}. 

\section{Conclusions}\label{sec:conclusions}

 \begin{table*}
\caption{Maximum likelihood parameters obtained with SPT data \citep{Reichardt2020} when RF predictions or templates are used for the SZ effects and templates for the other foregrounds. Priors are either Gaussian or flat. In particular, Gaussian priors corresponding to $1\sigma$ best-fit results from \citet{Planck2018} are imposed on cosmological parameters. The lower part of the table corresponds to derived parameters. Errors and upper limits correspond to $68\%$ and $95\%$ confidence intervals, respectively. Spectrum amplitudes are given in $\mu$K$^2$ and at $\ell=3000$. The last column presents the main results of this paper.}
\label{tab:results}      
\centering          
\begin{tabular}{l | c | c c c} 
\multirow{2}{*}{Parameter} & \multirow{2}{*}{Priors} & kSZ and tSZ & Cosmology-dependent \\ 
 & & templates & kSZ and tSZ \\
\hline  \hline
$\Omega_{\mathrm{b}}h^2$ & $0.0224\pm 0.0001$ & id. & id. & \\
$\Omega_{\mathrm{c}}h^2$ & $0.120 \pm 0.001$ & id. & id. & \\
$100 \theta_\mathrm{MC}$ & $1.0409 \pm 0.0003$ & id. & id. & \\
$n_s$ & $0.965\pm 0.004$ & id. & id. & \\
$\ln(10^{10}A_s)$ & 2.97 -- 3.12 & $3.026^{+0.022}_{-0.028}$ & $3.021^{+0.026}_{-0.028}$ & \\
$1-b$ & 0.5 -- 1.0 & -- & $0.57\pm 0.03$ & \\
$\zre$ & 5 -- 10 & $8.4^{+0.8}_{-1.3}$ & $7.9^{+1.1}_{-1.3}$ &  \\
d$z$ & 0 -- 4 & $1.8^{+1.3}_{-0.8}$  & $1.4^{+1.2}_{-0.6}$ &  \\
log~$[\alpha_0/\mathrm{Mpc}^3]$ & 2.5 -- 4.5 & -- &  $<4.1$  & \\
$\kappa$ [Mpc$^{-1}$] &  0.04 -- 0.20 & -- &  N.A. & \\
\hline
$\tau$ & -- & $0.068^{+0.009}_{-0.014}$ & $0.062^{+0.012}_{-0.015}$ &  \\
$\zend$ & $>4.5$ & $<8.6$ & $<8.5$ & \\
$\mathcal{D}^\mathrm{tSZ}$ & -- & $3.7\pm 0.5$  & $3.5\pm0.4$  & \\
$\mathcal{D}^\mathrm{kSZ}$ & -- & $3.1\pm 1.0$ & $3.4^{+0.5}_{-0.3}$ & \\
\hspace{2mm} patchy & -- & -- & $<1.6$ & \\
\hspace{2mm} late-time & -- & -- & $2.9^{+0.3}_{-0.2}$ & \\
\end{tabular}
\end{table*}

Using machine learning, we have included a cosmology-dependent derivation of the tSZ \citep[following, among others,][]{Komatsu:2002wc,2009MNRAS.392.1153T,2014A&A...571A..21P,Salvati2018} and kSZ angular power spectra, introduced in \citet{Gorce2020}, to the analysis of the SPT measurements of the small-scale power spectrum of CMB temperature fluctuations. Doing so, we have proven that the cosmological information included in the shape and amplitude of these foregrounds, usually considered as nuisance parameters, can help tighten constraints on cosmology and reionisation. In \citet{PAPER1}, we have focused our efforts on the tSZ effect. In this work, we have additionally included a proper modelling of reionisation and of the kSZ effect to the analysis.

Several works have demonstrated that the history of reionisation is not an instantaneous, redshift-symmetric process, motivating the inclusion of a physical parameterisation of reionisation \citep{Douspis2015} in the analysis of the latest \Planck~CMB data \citep{Planck2018}. In the framework introduced in this work, it is not the Thomson optical depth $\tau$ that is sampled, but the reionisation history itself, through two meaningful parameters: The reionisation midpoint $\zre$ and endpoint $\zend$, corresponding to the redshift when the Hydrogen in the IGM is $50\%$ and $100\%$ ionised, respectively. 
Analysing the \Planck~data, in Sec.~\ref{subsec:res_planck_asym}, we find a value of the Thomson optical depth $\tau=0.056\pm 0.07$ fully consistent with the one obtained using the fiducial, symmetric model ($\tau=0.054\pm0.007$). However, the reionisation history corresponding to this value gives a completely different picture of the high-redshift Universe, where the first light sources light up and start reionising the IGM as soon as $z=15$ (compared to $z=10$ for the instantaneous model, see Fig.~\ref{fig:res_large_scale}). Indeed, despite the $A_s$-$\tau$ degeneracy translating into a $A_s$-$\zre$ degeneracy, the \Planck~data alone gives tight constraints on the reionisation history, with a midpoint $\zre=7.4\pm 0.7$ and an upper limit on the end of the process $\zend<7.5$ (95\%~C.L.). These results corroborate previous works \citep[e.g.][]{Planck_2016_reio,QinPoulin_2020}, who found that the large-scale CMB data was really only sensitive to the integrated reionisation history, through $\tau$, and not to the details of the process. They motivate a new analysis of the small-scale CMB temperature power spectrum, designed to exploit the cosmological information included not only in the primary signal, but also in the CMB foregrounds, as done in Sec.~\ref{subsec:res_small_scale}.

In the analysis of the SPT small-scale data \citep{Reichardt2020}, we replace the template of the kSZ spectrum used in previous analyses by a cosmology-dependent estimate (Fig.~\ref{fig:ksz_template_vs_analytic}). Because a full derivation at each iteration of the sampler is too time-consuming to be considered, we train random forests to predict the patchy and late-time kSZ angular power spectra, when given the following parameters: $\Omega_b h^2$, $\Omega_c h^2$, $100\theta_\mathrm{MC}$, $n_s$, $A_s$, $\zre$, $\zend$, $\alpha_0$ and $\kappa$ -- the latter two defining the shape of the patchy kSZ spectrum \citep{Gorce2020}\footnote{The trained random forests are available at \url{https://szdb.osups.universite-paris-saclay.fr}.}. We incorporate the large-scale results to the small-scale analysis by imposing Gaussian priors corresponding to the results of Sec.~\ref{subsec:res_planck_asym} on $\Omega_b h^2$, $\Omega_c h^2$, $100\theta_\mathrm{MC}$ and $n_s$. We find that removing the tSZ and kSZ templates from the analysis, with or without letting the cosmology free, breaks the degeneracy between the amplitudes at $\ell=3000$ of the two power spectra. We report a $5\sigma$ measurement of the amplitude of the kSZ angular power spectrum, marginalised over cosmology (see Fig.~\ref{fig:triangle_atksz_free_cosmo}):
\[
\mathcal{D}_{3000}^\mathrm{kSZ}= 3.4^{+0.5}_{-0.3}~\mu\mathrm{K}^2\ (68\%~{\rm C.L.}).
\]
The previous measurement, obtained with the same data but fixed cosmology, was given at $3\sigma$ \citep{Reichardt2020}. We confirm their result of a negative tSZxCIB correlation, with a zero correlation being disfavoured above the $2\sigma$ level. 
These values result in a well-constrained reionisation history, although in slight tension with \Planck: the SPT data indeed seem to favour earlier reionisation scenarios, with (see Fig.~\ref{fig:triangle_reion_free_cosmo})
\[
\zre = 7.9^{+1.1}_{-1.3}\ (68\%~{\rm C.L.}),
\]
leading to an optical depth
\[\tau = 0.062^{+0.012}_{-0.015}\ (68\%~{\rm C.L.}).\] 
These scenarios are in agreement with current constraints on the IGM ionisation level obtained from Lyman-$\alpha$ emitters and quasar spectra (Fig.~\ref{fig:xe_spt}). We also report the first direct upper limit on the amplitude of the kSZ power stemming from reionisation 
\[ 
\mathcal{D}_{3000}^\mathrm{pkSZ}<1.6~\mu\mathrm{K}^2\ {\rm (95\% ~C.L.)}, 
\] corresponding to about half of the previous upper limit \citep{Reichardt2020} and show in Fig.~\ref{fig:sz_ps}, for the first time, the shape of the SZ spectra preferred by the small-scale CMB data. In particular, these new results seem to indicate a stronger small-scale power of both the tSZ and kSZ effects than previously allowed by templates.

If these results are promising, they can still be improved. In future work, we will focus our efforts on the modelling of other high-multipole CMB foregrounds, such as the cosmic infrared background \citep{ManiyarLagache_2019,ManiyarBethermin_2021}, and perform a simultaneous analysis of small- and large-scale CMB data in a unified framework to achieve better constraints on the reionisation history. We will also explore the potential of the future generation of CMB observatories to measure the shape of the kSZ angular power spectrum, and, doing so, to constrain the morphology of reionisation.

\begin{acknowledgements}
The authors thank Dominique Aubert, Ian Hothi, Stéphane Ilić and Adrian Liu for useful comments regarding a draft version of this paper. They additionally thank the referee for their useful comments which certainly helped improve this work.
AG's work was supported by the McGill Astrophysics Fellowship funded by the Trottier Chair in Astrophysics, as well as the Canadian Institute for Advanced Research (CIFAR) Azrieli Global Scholars program and the Canada 150 Programme. This work was additionally supported by the Programme National Cosmology et Galaxies (PNCG) of CNRS/INSU with INP and IN2P3, co-funded by CEA and CNES.

Most of the analysis was performed at the computation facility of IDOC\footnote{See \url{http://idoc.ias.u-psud.fr}.}, partly provided by DIM ACAV\footnote{See \url{http://www.dimacav-plus.fr}.}.
The authors also acknowledge the use of the supercomputer Cedar at Simon Fraser University, managed by Compute Canada to compute the training sets used in this work. The operation of this supercomputer is funded by the Canada Foundation for Innovation (CFI). \\

This research made use of \texttt{matplotlib}, a Python library for publication quality graphics \citep{hunter_2007}, of \texttt{scipy}, a Python-based ecosystem of open-source software for mathematics, science, and engineering \citep{scipy} -- including \texttt{numpy} \citep{numpy}, of the \texttt{scikit-learn} \citep{scikit-learn} library, and of the \texttt{corner} \citep{corner}, \texttt{getdist} \citep{Lewis_2019} and \texttt{arviz} \citep{arviz_2019} packages to analyse posterior distributions.

\end{acknowledgements}

\bibliographystyle{aa} 
\bibliography{biblio} 

\begin{thebibliography}{112}
\expandafter\ifx\csname natexlab\endcsname\relax\def\natexlab#1{#1}\fi

\bibitem[{{Abazajian} {et~al.}(2016){Abazajian}, {Adshead}, {Ahmed}, {Allen},
  {Alonso}, {Arnold}, {Baccigalupi}, {Bartlett}, {Battaglia}, {Benson},
  {Bischoff}, {Borrill}, {Buza}, {Calabrese}, {Caldwell}, {Carlstrom}, {Chang},
  {Crawford}, {Cyr-Racine}, {De Bernardis}, {de Haan}, {di Serego Alighieri},
  {Dunkley}, {Dvorkin}, {Errard}, {Fabbian}, {Feeney}, {Ferraro}, {Filippini},
  {Flauger}, {Fuller}, {Gluscevic}, {Green}, {Grin}, {Grohs}, {Henning},
  {Hill}, {Hlozek}, {Holder}, {Holzapfel}, {Hu}, {Huffenberger}, {Keskitalo},
  {Knox}, {Kosowsky}, {Kovac}, {Kovetz}, {Kuo}, {Kusaka}, {Le Jeune}, {Lee},
  {Lilley}, {Loverde}, {Madhavacheril}, {Mantz}, {Marsh}, {McMahon},
  {Meerburg}, {Meyers}, {Miller}, {Munoz}, {Nguyen}, {Niemack}, {Peloso},
  {Peloton}, {Pogosian}, {Pryke}, {Raveri}, {Reichardt}, {Rocha}, {Rotti},
  {Schaan}, {Schmittfull}, {Scott}, {Sehgal}, {Shandera}, {Sherwin}, {Smith},
  {Sorbo}, {Starkman}, {Story}, {van Engelen}, {Vieira}, {Watson}, {Whitehorn},
  \& {Kimmy Wu}}]{CMB-S4}
{Abazajian}, K.~N., {Adshead}, P., {Ahmed}, Z., {et~al.} 2016, arXiv e-prints,
  arXiv:1610.02743

\bibitem[{{Addison} {et~al.}(2012){Addison}, {Dunkley}, \&
  {Spergel}}]{AddisonDunkley_2012}
{Addison}, G.~E., {Dunkley}, J., \& {Spergel}, D.~N. 2012, \mnras, 427, 1741

\bibitem[{Ade {et~al.}(2019)Ade, Aguirre, Ahmed, Aiola, Ali,
  {et~al.}}]{Ade_2019}
Ade, P., Aguirre, J., Ahmed, Z., {et~al.} 2019, Journal of Cosmology and
  Astroparticle Physics, 2019, 056

\bibitem[{{Ahn} \& {Shapiro}(2021)}]{AhnShapiro_2021}
{Ahn}, K. \& {Shapiro}, P.~R. 2021, \apj, 914, 44

\bibitem[{{Alvarez}(2016)}]{Alvarez_2016}
{Alvarez}, M.~A. 2016, \apj, 824, 118

\bibitem[{{Alvarez} {et~al.}(2021){Alvarez}, {Ferraro}, {Hill}, {Hlo{\v{z}}ek},
  \& {Ikape}}]{AlvarezFerraro_2021}
{Alvarez}, M.~A., {Ferraro}, S., {Hill}, J.~C., {Hlo{\v{z}}ek}, R., \& {Ikape},
  M. 2021, \prd, 103, 063518

\bibitem[{An {et~al.}(1998)An, Brooks, \& Gelman}]{AnBrooksGelman_1998}
An, L., Brooks, S., \& Gelman, A. 1998, Journal of Computational and Graphical
  Statistics, 7, 434

\bibitem[{Aubert {et~al.}(2015)Aubert, Deparis, \& Ocvirk}]{aubert_2015_EMMA}
Aubert, D., Deparis, N., \& Ocvirk, P. 2015, \mnras, 454, 1012

\bibitem[{{Battaglia} {et~al.}(2013){Battaglia}, {Natarajan}, {Trac}, {Cen}, \&
  {Loeb}}]{battaglia_2013_paperIII}
{Battaglia}, N., {Natarajan}, A., {Trac}, H., {Cen}, R., \& {Loeb}, A. 2013,
  \apj, 776, 83

\bibitem[{{Baxter} {et~al.}(2021){Baxter}, {Weinberger}, {Haehnelt},
  {Ir{\v{s}}i{\v{c}}}, {Kulkarni}, {Pandey}, \& {Roy}}]{BaxterWeinberger_2021}
{Baxter}, E.~J., {Weinberger}, L., {Haehnelt}, M., {et~al.} 2021, \mnras, 501,
  6215

\bibitem[{{Becker} {et~al.}(2015){Becker}, {Bolton}, \&
  {Lidz}}]{BeckerBolton_2015}
{Becker}, G.~D., {Bolton}, J.~S., \& {Lidz}, A. 2015, \pasa, 32, e045

\bibitem[{{B{\'e}gin} {et~al.}(2022){B{\'e}gin}, {Liu}, \&
  {Gorce}}]{BeginLiu_2021}
{B{\'e}gin}, J.-M., {Liu}, A., \& {Gorce}, A. 2022, \prd, 105, 083503

\bibitem[{{Bennett} {et~al.}(1996){Bennett}, {Banday}, {Gorski}, {Hinshaw},
  {Jackson}, {Keegstra}, {Kogut}, {Smoot}, {Wilkinson}, \& {Wright}}]{COBE}
{Bennett}, C.~L., {Banday}, A.~J., {Gorski}, K.~M., {et~al.} 1996, \apjl, 464,
  L1

\bibitem[{{Billings} {et~al.}(2021){Billings}, {La Plante}, \&
  {Aguirre}}]{BillingsLaPlante_2021}
{Billings}, T.~S., {La Plante}, P., \& {Aguirre}, J.~E. 2021, arXiv e-prints,
  arXiv:2103.14563

\bibitem[{{Bosman} {et~al.}(2018){Bosman}, {Fan}, {Jiang}, {Reed}, {Matsuoka},
  {Becker}, \& {Haehnelt}}]{BosmanFan_2018}
{Bosman}, S. E.~I., {Fan}, X., {Jiang}, L., {et~al.} 2018, \mnras, 479, 1055

\bibitem[{{Bouwens} {et~al.}(2015){Bouwens}, {Illingworth}, {Oesch}, {Caruana},
  {Holwerda}, {Smit}, \& {Wilkins}}]{Bouwens2015}
{Bouwens}, R.~J., {Illingworth}, G.~D., {Oesch}, P.~A., {et~al.} 2015, \apj,
  811, 140

\bibitem[{{Calabrese} {et~al.}(2014){Calabrese}, {Hlo{\v{z}}ek}, {Battaglia},
  {Bond}, {de Bernardis}, {Devlin}, {Hajian}, {Henderson}, {Hil}, {Kosowsky},
  {Louis}, {McMahon}, {Moodley}, {Newburgh}, {Niemack}, {Page}, {Partridge},
  {Sehgal}, {Sievers}, {Spergel}, {Staggs}, {Switzer}, {Trac}, \&
  {Wollack}}]{CalabreseHlozek_2014}
{Calabrese}, E., {Hlo{\v{z}}ek}, R., {Battaglia}, N., {et~al.} 2014, \jcap,
  2014, 010

\bibitem[{{Casey} {et~al.}(2014){Casey}, {Narayanan}, \&
  {Cooray}}]{CaseyNarayanan_2014}
{Casey}, C.~M., {Narayanan}, D., \& {Cooray}, A. 2014, \physrep, 541, 45

\bibitem[{{Chatterjee} {et~al.}(2021){Chatterjee}, {Choudhury}, \&
  {Mitra}}]{ChatterjeeChoudhury_2021}
{Chatterjee}, A., {Choudhury}, T.~R., \& {Mitra}, S. 2021, \mnras, 507, 2405

\bibitem[{{Chen} {et~al.}(2022){Chen}, {Trac}, {Mukherjee}, \&
  {Cen}}]{ChenTrac_2022}
{Chen}, N., {Trac}, H., {Mukherjee}, S., \& {Cen}, R. 2022, arXiv e-prints,
  arXiv:2203.04337

\bibitem[{{Choudhury} {et~al.}(2021){Choudhury}, {Mukherjee}, \&
  {Paul}}]{ChoudhuryMukherjee_2021}
{Choudhury}, T.~R., {Mukherjee}, S., \& {Paul}, S. 2021, \mnras, 501, L7

\bibitem[{{Das} {et~al.}(2014){Das}, {Louis}, {Nolta}, {Addison},
  {Battistelli}, {Bond}, {Calabrese}, {Crichton}, {Devlin}, {Dicker},
  {Dunkley}, {D{\"u}nner}, {Fowler}, {Gralla}, {Hajian}, {Halpern},
  {Hasselfield}, {Hilton}, {Hincks}, {Hlozek}, {Huffenberger}, {Hughes},
  {Irwin}, {Kosowsky}, {Lupton}, {Marriage}, {Marsden}, {Menanteau}, {Moodley},
  {Niemack}, {Page}, {Partridge}, {Reese}, {Schmitt}, {Sehgal}, {Sherwin},
  {Sievers}, {Spergel}, {Staggs}, {Swetz}, {Switzer}, {Thornton}, {Trac}, \&
  {Wollack}}]{DasLouis_2014}
{Das}, S., {Louis}, T., {Nolta}, M.~R., {et~al.} 2014, \jcap, 2014, 014

\bibitem[{{Douspis} {et~al.}(2015){Douspis}, {Aghanim}, {Ili{\'c}}, \&
  {Langer}}]{Douspis2015}
{Douspis}, M., {Aghanim}, N., {Ili{\'c}}, S., \& {Langer}, M. 2015, \aap, 580,
  L4

\bibitem[{{Douspis} {et~al.}(2006){Douspis}, {Aghanim}, \&
  {Langer}}]{2006A&A...456..819D}
{Douspis}, M., {Aghanim}, N., \& {Langer}, M. 2006, \aap, 456, 819

\bibitem[{{Douspis} {et~al.}(2022){Douspis}, {Salvati}, {Gorce}, \&
  {Aghanim}}]{PAPER1}
{Douspis}, M., {Salvati}, L., {Gorce}, A., \& {Aghanim}, N. 2022, \aap, 659,
  A99

\bibitem[{{Dunkley} {et~al.}(2013){Dunkley}, {Calabrese}, {Sievers}, {Addison},
  {Battaglia}, {Battistelli}, {Bond}, {Das}, {Devlin}, {D{\"u}nner}, {Fowler},
  {Gralla}, {Hajian}, {Halpern}, {Hasselfield}, {Hincks}, {Hlozek}, {Hughes},
  {Irwin}, {Kosowsky}, {Louis}, {Marriage}, {Marsden}, {Menanteau}, {Moodley},
  {Niemack}, {Nolta}, {Page}, {Partridge}, {Sehgal}, {Spergel}, {Staggs},
  {Switzer}, {Trac}, \& {Wollack}}]{DunkleyCalabrese_2013}
{Dunkley}, J., {Calabrese}, E., {Sievers}, J., {et~al.} 2013, \jcap, 2013, 025

\bibitem[{{Dvorkin} \& {Smith}(2009)}]{DvorkinSmith_2009}
{Dvorkin}, C. \& {Smith}, K.~M. 2009, \prd, 79, 043003

\bibitem[{Foreman-Mackey(2016)}]{corner}
Foreman-Mackey, D. 2016, The Journal of Open Source Software, 1, 24

\bibitem[{{Galli} {et~al.}(2014){Galli}, {Benabed}, {Bouchet}, {Cardoso},
  {Elsner}, {Hivon}, {Mangilli}, {Prunet}, \& {Wandelt}}]{GalliBenabed_2014}
{Galli}, S., {Benabed}, K., {Bouchet}, F., {et~al.} 2014, \prd, 90, 063504

\bibitem[{{Gelman} \& {Rubin}(1992)}]{GelmanRubin_1992}
{Gelman}, A. \& {Rubin}, D.~B. 1992, Statistical Science, 7, 457

\bibitem[{{George} {et~al.}(2015){George}, {Reichardt}, {Aird}, {Benson},
  {Bleem}, {Carlstrom}, {Chang}, {Cho}, {Crawford}, {Crites}, {de Haan},
  {Dobbs}, {Dudley}, {Halverson}, {Harrington}, {Holder}, {Holzapfel}, {Hou},
  {Hrubes}, {Keisler}, {Knox}, {Lee}, {Leitch}, {Lueker}, {Luong-Van},
  {McMahon}, {Mehl}, {Meyer}, {Millea}, {Mocanu}, {Mohr}, {Montroy}, {Padin},
  {Plagge}, {Pryke}, {Ruhl}, {Schaffer}, {Shaw}, {Shirokoff}, {Spieler},
  {Staniszewski}, {Stark}, {Story}, {van Engelen}, {Vanderlinde}, {Vieira},
  {Williamson}, \& {Zahn}}]{GeorgeReichardt_2015}
{George}, E.~M., {Reichardt}, C.~L., {Aird}, K.~A., {et~al.} 2015, \apj, 799,
  177

\bibitem[{{Gluscevic} {et~al.}(2013){Gluscevic}, {Kamionkowski}, \&
  {Hanson}}]{GluscevicKamionkowski_2013}
{Gluscevic}, V., {Kamionkowski}, M., \& {Hanson}, D. 2013, \prd, 87, 047303

\bibitem[{{Gorce} {et~al.}(2018){Gorce}, {Douspis}, {Aghanim}, \&
  {Langer}}]{GorceDouspis_2018}
{Gorce}, A., {Douspis}, M., {Aghanim}, N., \& {Langer}, M. 2018, \aap, 616,
  A113

\bibitem[{{Gorce} {et~al.}(2020){Gorce}, {Ili{\'c}}, {Douspis}, {Aubert}, \&
  {Langer}}]{Gorce2020}
{Gorce}, A., {Ili{\'c}}, S., {Douspis}, M., {Aubert}, D., \& {Langer}, M. 2020,
  \aap, 640, A90

\bibitem[{{Greig} \& {Mesinger}(2015)}]{GreigMesinger_2015}
{Greig}, B. \& {Mesinger}, A. 2015, \mnras, 449, 4246

\bibitem[{{Greig} {et~al.}(2019){Greig}, {Mesinger}, \&
  {Ba{\~n}ados}}]{GreigMesinger_2019}
{Greig}, B., {Mesinger}, A., \& {Ba{\~n}ados}, E. 2019, \mnras, 484, 5094

\bibitem[{{Greig} {et~al.}(2017){Greig}, {Mesinger}, {Haiman}, \&
  {Simcoe}}]{GreigMesinger_2017}
{Greig}, B., {Mesinger}, A., {Haiman}, Z., \& {Simcoe}, R.~A. 2017, \mnras,
  466, 4239

\bibitem[{{Hoag} {et~al.}(2019){Hoag}, {Brada{\v{c}}}, {Huang}, {Mason},
  {Treu}, {Schmidt}, {Trenti}, {Strait}, {Lemaux}, {Finney}, \&
  {Paddock}}]{HoagBradac_2019}
{Hoag}, A., {Brada{\v{c}}}, M., {Huang}, K., {et~al.} 2019, \apj, 878, 12

\bibitem[{{Holder} {et~al.}(2007){Holder}, {Iliev}, \&
  {Mellema}}]{HolderIliev_2007}
{Holder}, G.~P., {Iliev}, I.~T., \& {Mellema}, G. 2007, \apjl, 663, L1

\bibitem[{Hotinli \& Johnson(2022)}]{HotinliJohnson_2020}
Hotinli, S.~C. \& Johnson, M.~C. 2022, Phys. Rev. D, 105, 063522

\bibitem[{Howlett {et~al.}(2012)Howlett, Lewis, Hall, \& Challinor}]{camb2}
Howlett, C., Lewis, A., Hall, A., \& Challinor, A. 2012, \jcap, 1204, 027

\bibitem[{{Hu} \& {Holder}(2003)}]{HuHolder_2003}
{Hu}, W. \& {Holder}, G.~P. 2003, \prd, 68, 023001

\bibitem[{Hunter(2007)}]{hunter_2007}
Hunter, J.~D. 2007, Computing In Science \& Engineering, 9, 90

\bibitem[{{Ichiki}(2014)}]{Ichiki_2014}
{Ichiki}, K. 2014, Progress of Theoretical and Experimental Physics, 2014,
  06B109

\bibitem[{{Iliev} {et~al.}(2007){Iliev}, {Pen}, {Bond}, {Mellema}, \&
  {Shapiro}}]{iliev_2007}
{Iliev}, I.~T., {Pen}, U.-L., {Bond}, J.~R., {Mellema}, G., \& {Shapiro}, P.~R.
  2007, \apj, 660, 933

\bibitem[{{Jones} {et~al.}(2001){Jones}, {Oliphant}, {Peterson},
  {et~al.}}]{scipy}
{Jones}, E., {Oliphant}, T., {Peterson}, P., {et~al.} 2001, {SciPy}: Open
  source scientific tools for {Python}

\bibitem[{{Jung} {et~al.}(2020){Jung}, {Finkelstein}, {Dickinson}, {Hutchison},
  {Larson}, {Papovich}, {Pentericci}, {Straughn}, {Guo}, {Malhotra}, {Rhoads},
  {Song}, {Tilvi}, \& {Wold}}]{JungFinkelstein_2020}
{Jung}, I., {Finkelstein}, S.~L., {Dickinson}, M., {et~al.} 2020, \apj, 904,
  144

\bibitem[{Komatsu \& Seljak(2002)}]{Komatsu:2002wc}
Komatsu, E. \& Seljak, U. 2002, Mon. Not. Roy. Astron. Soc., 336, 1256

\bibitem[{{Konno} {et~al.}(2018){Konno}, {Ouchi}, {Shibuya}, {Ono},
  {Shimasaku}, {Taniguchi}, {Nagao}, {Kobayashi}, {Kajisawa}, {Kashikawa},
  {Inoue}, {Oguri}, {Furusawa}, {Goto}, {Harikane}, {Higuchi}, {Komiyama},
  {Kusakabe}, {Miyazaki}, {Nakajima}, \& {Wang}}]{konno_2017}
{Konno}, A., {Ouchi}, M., {Shibuya}, T., {et~al.} 2018, \pasj, 70, S16

\bibitem[{{Kosowsky}(2003)}]{ACT}
{Kosowsky}, A. 2003, \nar, 47, 939

\bibitem[{{Kuhlen} \& {Faucher-Gigu{\`e}re}(2012)}]{kuhlen_2012}
{Kuhlen}, M. \& {Faucher-Gigu{\`e}re}, C.-A. 2012, \mnras, 423, 862

\bibitem[{Kumar {et~al.}(2019)Kumar, Carroll, Hartikainen, \&
  Martin}]{arviz_2019}
Kumar, R., Carroll, C., Hartikainen, A., \& Martin, O. 2019, Journal of Open
  Source Software, 4, 1143

\bibitem[{{La Plante} {et~al.}(2020){La Plante}, {Lidz}, {Aguirre}, \&
  {Kohn}}]{LaPlanteLidz_2020}
{La Plante}, P., {Lidz}, A., {Aguirre}, J., \& {Kohn}, S. 2020, \apj, 899, 40

\bibitem[{{La Plante} {et~al.}(2022){La Plante}, {Sipple}, \&
  {Lidz}}]{LaPlanteSipple_2022}
{La Plante}, P., {Sipple}, J., \& {Lidz}, A. 2022, \apj, 928, 162

\bibitem[{{Lagache} {et~al.}(2005){Lagache}, {Puget}, \&
  {Dole}}]{LagachePuget_2005}
{Lagache}, G., {Puget}, J.-L., \& {Dole}, H. 2005, \araa, 43, 727

\bibitem[{Lewis(2013)}]{cosmosmc2}
Lewis, A. 2013, \prd, 87, 103529

\bibitem[{Lewis(2019)}]{Lewis_2019}
Lewis, A. 2019 [\eprint[arXiv]{1910.13970}]

\bibitem[{Lewis \& Bridle(2002)}]{lewis_2002_cosmomc}
Lewis, A. \& Bridle, S. 2002, Phys. Rev., D66, 103511

\bibitem[{Lewis {et~al.}(2000)Lewis, Challinor, \& Lasenby}]{camb1}
Lewis, A., Challinor, A., \& Lasenby, A. 2000, \apj, 538, 473

\bibitem[{{Liu} {et~al.}(2016){Liu}, {Pritchard}, {Allison}, {Parsons},
  {Seljak}, \& {Sherwin}}]{LiuPritchard_2016}
{Liu}, A., {Pritchard}, J.~R., {Allison}, R., {et~al.} 2016, \prd, 93, 043013

\bibitem[{{Ma} {et~al.}(2018){Ma}, {Helgason}, {Komatsu}, {Ciardi}, \&
  {Ferrara}}]{MaHelgason_2018}
{Ma}, Q., {Helgason}, K., {Komatsu}, E., {Ciardi}, B., \& {Ferrara}, A. 2018,
  \mnras, 476, 4025

\bibitem[{{Maniyar} {et~al.}(2021){Maniyar}, {B{\'e}thermin}, \&
  {Lagache}}]{ManiyarBethermin_2021}
{Maniyar}, A., {B{\'e}thermin}, M., \& {Lagache}, G. 2021, \aap, 645, A40

\bibitem[{{Maniyar} {et~al.}(2019){Maniyar}, {Lagache}, {B{\'e}thermin}, \&
  {Ili{\'c}}}]{ManiyarLagache_2019}
{Maniyar}, A., {Lagache}, G., {B{\'e}thermin}, M., \& {Ili{\'c}}, S. 2019,
  \aap, 621, A32

\bibitem[{{Mason} {et~al.}(2018){Mason}, {Treu}, {Dijkstra}, {Mesinger},
  {Trenti}, {Pentericci}, {de Barros}, \& {Vanzella}}]{mason_2017}
{Mason}, C.~A., {Treu}, T., {Dijkstra}, M., {et~al.} 2018, \apj, 856, 2

\bibitem[{{McGreer} {et~al.}(2015){McGreer}, {Mesinger}, \&
  {D'Odorico}}]{mcgreer_2015}
{McGreer}, I.~D., {Mesinger}, A., \& {D'Odorico}, V. 2015, \mnras, 447, 499

\bibitem[{{McQuinn} {et~al.}(2005){McQuinn}, {Furlanetto}, {Hernquist}, {Zahn},
  \& {Zaldarriaga}}]{mcquinn_2005}
{McQuinn}, M., {Furlanetto}, S.~R., {Hernquist}, L., {Zahn}, O., \&
  {Zaldarriaga}, M. 2005, \apj, 630, 643

\bibitem[{{Mellema} {et~al.}(2015){Mellema}, {Koopmans}, {Shukla}, {Datta},
  {Mesinger}, \& {Majumdar}}]{MellemaKoopmans_2015}
{Mellema}, G., {Koopmans}, L., {Shukla}, H., {et~al.} 2015, in Advancing
  Astrophysics with the Square Kilometre Array (AASKA14), 10

\bibitem[{{Mesinger} {et~al.}(2011){Mesinger}, {Furlanetto}, \&
  {Cen}}]{21cmFAST_2011}
{Mesinger}, A., {Furlanetto}, S., \& {Cen}, R. 2011, \mnras, 411, 955

\bibitem[{Mesinger {et~al.}(2012)Mesinger, McQuinn, \&
  Spergel}]{mesinger_2012_kSZ}
Mesinger, A., McQuinn, M., \& Spergel, D.~N. 2012, \mnras, 422, 1403

\bibitem[{{Millea} \& {Bouchet}(2018)}]{MilleaBouchet_2018}
{Millea}, M. \& {Bouchet}, F. 2018, \aap, 617, A96

\bibitem[{{Mortonson} \& {Hu}(2008)}]{MortonsonHu_2008}
{Mortonson}, M.~J. \& {Hu}, W. 2008, \apj, 672, 737

\bibitem[{Murray {et~al.}(2020)Murray, Greig, Mesinger, Muñoz, Qin, Park, \&
  Watkinson}]{Murray2020}
Murray, S.~G., Greig, B., Mesinger, A., {et~al.} 2020, Journal of Open Source
  Software, 5, 2582

\bibitem[{{Namikawa} {et~al.}(2021){Namikawa}, {Roy}, {Sherwin}, {Battaglia},
  \& {Spergel}}]{NamikawaRoy_2021}
{Namikawa}, T., {Roy}, A., {Sherwin}, B.~D., {Battaglia}, N., \& {Spergel},
  D.~N. 2021, \prd, 104, 063514

\bibitem[{{Natale} {et~al.}(2020){Natale}, {Pagano}, {Lattanzi}, {Migliaccio},
  {Colombo}, {Gruppuso}, {Natoli}, \& {Polenta}}]{NatalePagano_2020}
{Natale}, U., {Pagano}, L., {Lattanzi}, M., {et~al.} 2020, \aap, 644, A32

\bibitem[{Oliphant(2006)}]{numpy}
Oliphant, T. 2006, {NumPy}: A guide to {NumPy}, USA: Trelgol Publishing,
  [Online; accessed <today>]

\bibitem[{Ota {et~al.}(2008)Ota, Iye, Kashikawa, Shimasaku, Kobayashi, Totani,
  Nagashima, Morokuma, Furusawa, Hattori, Matsuda, Hashimoto, \&
  Ouchi}]{ota_2008}
Ota, K., Iye, M., Kashikawa, N., {et~al.} 2008, ApJ, 677, 12

\bibitem[{{Pagano} {et~al.}(2020){Pagano}, {Delouis}, {Mottet}, {Puget}, \&
  {Vibert}}]{PaganoDelouis_2020}
{Pagano}, L., {Delouis}, J.~M., {Mottet}, S., {Puget}, J.~L., \& {Vibert}, L.
  2020, \aap, 635, A99

\bibitem[{{Park} {et~al.}(2013){Park}, {Shapiro}, {Komatsu}, {Iliev}, {Ahn}, \&
  {Mellema}}]{ParkShapiro_2013}
{Park}, H., {Shapiro}, P.~R., {Komatsu}, E., {et~al.} 2013, \apj, 769, 93

\bibitem[{{Park} {et~al.}(2019){Park}, {Mesinger}, {Greig}, \&
  {Gillet}}]{ParkMesinger_2019}
{Park}, J., {Mesinger}, A., {Greig}, B., \& {Gillet}, N. 2019, \mnras, 484, 933

\bibitem[{{Paul} {et~al.}(2021){Paul}, {Mukherjee}, \&
  {Choudhury}}]{PaulMukherjee_2021}
{Paul}, S., {Mukherjee}, S., \& {Choudhury}, T.~R. 2021, \mnras, 500, 232

\bibitem[{Pedregosa {et~al.}(2011)Pedregosa, Varoquaux, Gramfort, Michel,
  Thirion, Grisel, Blondel, Prettenhofer, Weiss, Dubourg, Vanderplas, Passos,
  Cournapeau, Brucher, Perrot, \& Duchesnay}]{scikit-learn}
Pedregosa, F., Varoquaux, G., Gramfort, A., {et~al.} 2011, Journal of Machine
  Learning Research, 12, 2825

\bibitem[{{Planck Collaboration} {et~al.}(2016{\natexlab{a}}){Planck
  Collaboration}, {Adam}, {Ade}, {Aghanim}, {Akrami}, {Alves}, {Arg{\"u}eso},
  {Arnaud}, {Arroja}, {Ashdown}, \& et~al.}]{PlanckCollaborationAdam_2016}
{Planck Collaboration}, {Adam}, R., {Ade}, P.~A.~R., {et~al.}
  2016{\natexlab{a}}, \aap, 594, A1

\bibitem[{{Planck Collaboration} {et~al.}(2016{\natexlab{b}}){Planck
  Collaboration}, {Adam}, {Aghanim}, {Ashdown}, {Aumont}, {Baccigalupi},
  {Ballardini}, {Banday}, {Barreiro}, {Bartolo}, {Basak}, {Battye}, {Benabed},
  {Bernard}, {Bersanelli}, {Bielewicz}, {Bock}, {Bonaldi}, {Bonavera}, {Bond},
  {Borrill}, {Bouchet}, {Boulanger}, {Bucher}, {Burigana}, {Calabrese},
  {Cardoso}, {Carron}, {Chiang}, {Colombo}, {Combet}, {Comis}, {Couchot},
  {Coulais}, {Crill}, {Curto}, {Cuttaia}, {Davis}, {de Bernardis}, {de Rosa},
  {de Zotti}, {Delabrouille}, {Di Valentino}, {Dickinson}, {Diego}, {Dor{\'e}},
  {Douspis}, {Ducout}, {Dupac}, {Elsner}, {En{\ss}lin}, {Eriksen}, {Falgarone},
  {Fantaye}, {Finelli}, {Forastieri}, {Frailis}, {Fraisse}, {Franceschi},
  {Frolov}, {Galeotta}, {Galli}, {Ganga}, {G{\'e}nova-Santos}, {Gerbino},
  {Ghosh}, {Gonz{\'a}lez-Nuevo}, {G{\'o}rski}, {Gruppuso}, {Gudmundsson},
  {Hansen}, {Helou}, {Henrot-Versill{\'e}}, {Herranz}, {Hivon}, {Huang},
  {Ili{\'c}}, {Jaffe}, {Jones}, {Keih{\"a}nen}, {Keskitalo}, {Kisner}, {Knox},
  {Krachmalnicoff}, {Kunz}, {Kurki-Suonio}, {Lagache}, {L{\"a}hteenm{\"a}ki},
  {Lamarre}, {Langer}, {Lasenby}, {Lattanzi}, {Lawrence}, {Le Jeune},
  {Levrier}, {Lewis}, {Liguori}, {Lilje}, {L{\'o}pez-Caniego}, {Ma},
  {Mac{\'\i}as-P{\'e}rez}, {Maggio}, {Mangilli}, {Maris}, {Martin},
  {Mart{\'\i}nez-Gonz{\'a}lez}, {Matarrese}, {Mauri}, {McEwen}, {Meinhold},
  {Melchiorri}, {Mennella}, {Migliaccio}, {Miville-Desch{\^e}nes}, {Molinari},
  {Moneti}, {Montier}, {Morgante}, {Moss}, {Naselsky}, {Natoli}, {Oxborrow},
  {Pagano}, {Paoletti}, {Partridge}, {Patanchon}, {Patrizii}, {Perdereau},
  {Perotto}, {Pettorino}, {Piacentini}, {Plaszczynski}, {Polastri}, {Polenta},
  {Puget}, {Rachen}, {Racine}, {Reinecke}, {Remazeilles}, {Renzi}, {Rocha},
  {Rossetti}, {Roudier}, {Rubi{\~n}o-Mart{\'\i}n}, {Ruiz-Granados}, {Salvati},
  {Sandri}, {Savelainen}, {Scott}, {Sirri}, {Sunyaev}, {Suur-Uski}, {Tauber},
  {Tenti}, {Toffolatti}, {Tomasi}, {Tristram}, {Trombetti}, {Valiviita}, {Van
  Tent}, {Vielva}, {Villa}, {Vittorio}, {Wandelt}, {Wehus}, {White}, {Zacchei},
  \& {Zonca}}]{Planck_2016_reio}
{Planck Collaboration}, {Adam}, R., {Aghanim}, N., {et~al.} 2016{\natexlab{b}},
  \aap, 596, A108

\bibitem[{{Planck Collaboration} {et~al.}(2014){Planck Collaboration}, {Ade},
  {Aghanim}, {Armitage-Caplan}, {Arnaud}, {Ashdown}, {Atrio-Barandela},
  {Aumont}, {Baccigalupi}, {Banday}, {Barreiro}, {Bartlett}, {Battaner},
  {Benabed}, {Beno{\^\i}t}, {Benoit-L{\'e}vy}, {Bernard}, {Bersanelli},
  {Bielewicz}, {Bobin}, {Bock}, {Bonaldi}, {Bond}, {Borrill}, {Bouchet},
  {Bridges}, {Bucher}, {Burigana}, {Butler}, {Cardoso}, {Carvalho}, {Catalano},
  {Challinor}, {Chamballu}, {Chiang}, {Chiang}, {Christensen}, {Church},
  {Clements}, {Colombi}, {Colombo}, {Comis}, {Couchot}, {Coulais}, {Crill},
  {Curto}, {Cuttaia}, {Da Silva}, {Danese}, {Davies}, {Davis}, {de Bernardis},
  {de Rosa}, {de Zotti}, {Delabrouille}, {Delouis}, {D{\'e}sert}, {Dickinson},
  {Diego}, {Dolag}, {Dole}, {Donzelli}, {Dor{\'e}}, {Douspis}, {Dupac},
  {Efstathiou}, {En{\ss}lin}, {Eriksen}, {Finelli}, {Flores-Cacho}, {Forni},
  {Frailis}, {Franceschi}, {Galeotta}, {Ganga}, {G{\'e}nova-Santos}, {Giard},
  {Giardino}, {Giraud-H{\'e}raud}, {Gonz{\'a}lez-Nuevo}, {G{\'o}rski},
  {Gratton}, {Gregorio}, {Gruppuso}, {Hansen}, {Hanson}, {Harrison},
  {Henrot-Versill{\'e}}, {Hern{\'a}ndez-Monteagudo}, {Herranz}, {Hildebrandt},
  {Hivon}, {Hobson}, {Holmes}, {Hornstrup}, {Hovest}, {Huffenberger}, {Hurier},
  {Jaffe}, {Jaffe}, {Jones}, {Juvela}, {Keih{\"a}nen}, {Keskitalo}, {Kisner},
  {Kneissl}, {Knoche}, {Knox}, {Kunz}, {Kurki-Suonio}, {Lacasa}, {Lagache},
  {L{\"a}hteenm{\"a}ki}, {Lamarre}, {Lasenby}, {Laureijs}, {Lawrence}, {Leahy},
  {Leonardi}, {Le{\'o}n-Tavares}, {Lesgourgues}, {Liguori}, {Lilje},
  {Linden-V{\o}rnle}, {L{\'o}pez-Caniego}, {Lubin}, {Mac{\'\i}as-P{\'e}rez},
  {Maffei}, {Maino}, {Mandolesi}, {Marcos-Caballero}, {Maris}, {Marshall},
  {Martin}, {Mart{\'\i}nez-Gonz{\'a}lez}, {Masi}, {Massardi}, {Matarrese},
  {Matthai}, {Mazzotta}, {Melchiorri}, {Melin}, {Mendes}, {Mennella},
  {Migliaccio}, {Mitra}, {Miville-Desch{\^e}nes}, {Moneti}, {Montier},
  {Morgante}, {Mortlock}, {Moss}, {Munshi}, {Naselsky}, {Nati}, {Natoli},
  {Netterfield}, {N{\o}rgaard-Nielsen}, {Noviello}, {Novikov}, {Novikov},
  {Osborne}, {Oxborrow}, {Paci}, {Pagano}, {Pajot}, {Paoletti}, {Partridge},
  {Pasian}, {Patanchon}, {Perdereau}, {Perotto}, {Perrotta}, {Piacentini},
  {Piat}, {Pierpaoli}, {Pietrobon}, {Plaszczynski}, {Pointecouteau}, {Polenta},
  {Ponthieu}, {Popa}, {Poutanen}, {Pratt}, {Pr{\'e}zeau}, {Prunet}, {Puget},
  {Rachen}, {Rebolo}, {Reinecke}, {Remazeilles}, {Renault}, {Ricciardi},
  {Riller}, {Ristorcelli}, {Rocha}, {Rosset}, {Rossetti}, {Roudier},
  {Rubi{\~n}o-Mart{\'\i}n}, {Rusholme}, {Sandri}, {Santos}, {Savini}, {Scott},
  {Seiffert}, {Shellard}, {Spencer}, {Starck}, {Stolyarov}, {Stompor},
  {Sudiwala}, {Sunyaev}, {Sureau}, {Sutton}, {Suur-Uski}, {Sygnet}, {Tauber},
  {Tavagnacco}, {Terenzi}, {Toffolatti}, {Tomasi}, {Tristram}, {Tucci},
  {Tuovinen}, {Umana}, {Valenziano}, {Valiviita}, {Van Tent}, {Varis},
  {Vielva}, {Villa}, {Vittorio}, {Wade}, {Wandelt}, {White}, {Yvon}, {Zacchei},
  \& {Zonca}}]{2014A&A...571A..21P}
{Planck Collaboration}, {Ade}, P.~A.~R., {Aghanim}, N., {et~al.} 2014, \aap,
  571, A21

\bibitem[{{Planck Collaboration} {et~al.}(2020{\natexlab{a}}){Planck
  Collaboration}, {Aghanim}, {Akrami}, {Arroja}, {Ashdown}, {Aumont},
  {Baccigalupi}, {Ballardini}, {Banday}, {Barreiro}, {Bartolo}, {Basak},
  {Battye}, {Benabed}, {Bernard}, {Bersanelli}, {Bielewicz}, {Bock}, {Bond},
  {Borrill}, {Bouchet}, {Boulanger}, {Bucher}, {Burigana}, {Butler},
  {Calabrese}, {Cardoso}, {Carron}, {Casaponsa}, {Challinor}, {Chiang},
  {Colombo}, {Combet}, {Contreras}, {Crill}, {Cuttaia}, {de Bernardis}, {de
  Zotti}, {Delabrouille}, {Delouis}, {D{\'e}sert}, {Di Valentino}, {Dickinson},
  {Diego}, {Donzelli}, {Dor{\'e}}, {Douspis}, {Ducout}, {Dupac}, {Efstathiou},
  {Elsner}, {En{\ss}lin}, {Eriksen}, {Falgarone}, {Fantaye}, {Fergusson},
  {Fernandez-Cobos}, {Finelli}, {Forastieri}, {Frailis}, {Franceschi},
  {Frolov}, {Galeotta}, {Galli}, {Ganga}, {G{\'e}nova-Santos}, {Gerbino},
  {Ghosh}, {Gonz{\'a}lez-Nuevo}, {G{\'o}rski}, {Gratton}, {Gruppuso},
  {Gudmundsson}, {Hamann}, {Handley}, {Hansen}, {Helou}, {Herranz},
  {Hildebrandt}, {Hivon}, {Huang}, {Jaffe}, {Jones}, {Karakci}, {Keih{\"a}nen},
  {Keskitalo}, {Kiiveri}, {Kim}, {Kisner}, {Knox}, {Krachmalnicoff}, {Kunz},
  {Kurki-Suonio}, {Lagache}, {Lamarre}, {Langer}, {Lasenby}, {Lattanzi},
  {Lawrence}, {Le Jeune}, {Leahy}, {Lesgourgues}, {Levrier}, {Lewis},
  {Liguori}, {Lilje}, {Lilley}, {Lindholm}, {L{\'o}pez-Caniego}, {Lubin}, {Ma},
  {Mac{\'\i}as-P{\'e}rez}, {Maggio}, {Maino}, {Mandolesi}, {Mangilli},
  {Marcos-Caballero}, {Maris}, {Martin}, {Martinelli},
  {Mart{\'\i}nez-Gonz{\'a}lez}, {Matarrese}, {Mauri}, {McEwen}, {Meerburg},
  {Meinhold}, {Melchiorri}, {Mennella}, {Migliaccio}, {Millea}, {Mitra},
  {Miville-Desch{\^e}nes}, {Molinari}, {Moneti}, {Montier}, {Morgante}, {Moss},
  {Mottet}, {M{\"u}nchmeyer}, {Natoli}, {N{\o}rgaard-Nielsen}, {Oxborrow},
  {Pagano}, {Paoletti}, {Partridge}, {Patanchon}, {Pearson}, {Peel}, {Peiris},
  {Perrotta}, {Pettorino}, {Piacentini}, {Polastri}, {Polenta}, {Puget},
  {Rachen}, {Reinecke}, {Remazeilles}, {Renault}, {Renzi}, {Rocha}, {Rosset},
  {Roudier}, {Rubi{\~n}o-Mart{\'\i}n}, {Ruiz-Granados}, {Salvati}, {Sandri},
  {Savelainen}, {Scott}, {Shellard}, {Shiraishi}, {Sirignano}, {Sirri},
  {Spencer}, {Sunyaev}, {Suur-Uski}, {Tauber}, {Tavagnacco}, {Tenti},
  {Terenzi}, {Toffolatti}, {Tomasi}, {Trombetti}, {Valiviita}, {Van Tent},
  {Vibert}, {Vielva}, {Villa}, {Vittorio}, {Wandelt}, {Wehus}, {White},
  {White}, {Zacchei}, \& {Zonca}}]{Planck2018}
{Planck Collaboration}, {Aghanim}, N., {Akrami}, Y., {et~al.}
  2020{\natexlab{a}}, \aap, 641, A1

\bibitem[{{Planck Collaboration} {et~al.}(2020{\natexlab{b}}){Planck
  Collaboration}, {Aghanim}, {Akrami}, {Ashdown}, {Aumont}, {Baccigalupi},
  {Ballardini}, {Banday}, {Barreiro}, {Bartolo}, {Basak}, {Battye}, {Benabed},
  {Bernard}, {Bersanelli}, {Bielewicz}, {Bock}, {Bond}, {Borrill}, {Bouchet},
  {Boulanger}, {Bucher}, {Burigana}, {Butler}, {Calabrese}, {Cardoso},
  {Carron}, {Challinor}, {Chiang}, {Chluba}, {Colombo}, {Combet}, {Contreras},
  {Crill}, {Cuttaia}, {de Bernardis}, {de Zotti}, {Delabrouille}, {Delouis},
  {Di Valentino}, {Diego}, {Dor{\'e}}, {Douspis}, {Ducout}, {Dupac}, {Dusini},
  {Efstathiou}, {Elsner}, {En{\ss}lin}, {Eriksen}, {Fantaye}, {Farhang},
  {Fergusson}, {Fernandez-Cobos}, {Finelli}, {Forastieri}, {Frailis},
  {Fraisse}, {Franceschi}, {Frolov}, {Galeotta}, {Galli}, {Ganga},
  {G{\'e}nova-Santos}, {Gerbino}, {Ghosh}, {Gonz{\'a}lez-Nuevo}, {G{\'o}rski},
  {Gratton}, {Gruppuso}, {Gudmundsson}, {Hamann}, {Handley}, {Hansen},
  {Herranz}, {Hildebrandt}, {Hivon}, {Huang}, {Jaffe}, {Jones}, {Karakci},
  {Keih{\"a}nen}, {Keskitalo}, {Kiiveri}, {Kim}, {Kisner}, {Knox},
  {Krachmalnicoff}, {Kunz}, {Kurki-Suonio}, {Lagache}, {Lamarre}, {Lasenby},
  {Lattanzi}, {Lawrence}, {Le Jeune}, {Lemos}, {Lesgourgues}, {Levrier},
  {Lewis}, {Liguori}, {Lilje}, {Lilley}, {Lindholm}, {L{\'o}pez-Caniego},
  {Lubin}, {Ma}, {Mac{\'\i}as-P{\'e}rez}, {Maggio}, {Maino}, {Mandolesi},
  {Mangilli}, {Marcos-Caballero}, {Maris}, {Martin}, {Martinelli},
  {Mart{\'\i}nez-Gonz{\'a}lez}, {Matarrese}, {Mauri}, {McEwen}, {Meinhold},
  {Melchiorri}, {Mennella}, {Migliaccio}, {Millea}, {Mitra},
  {Miville-Desch{\^e}nes}, {Molinari}, {Montier}, {Morgante}, {Moss}, {Natoli},
  {N{\o}rgaard-Nielsen}, {Pagano}, {Paoletti}, {Partridge}, {Patanchon},
  {Peiris}, {Perrotta}, {Pettorino}, {Piacentini}, {Polastri}, {Polenta},
  {Puget}, {Rachen}, {Reinecke}, {Remazeilles}, {Renzi}, {Rocha}, {Rosset},
  {Roudier}, {Rubi{\~n}o-Mart{\'\i}n}, {Ruiz-Granados}, {Salvati}, {Sandri},
  {Savelainen}, {Scott}, {Shellard}, {Sirignano}, {Sirri}, {Spencer},
  {Sunyaev}, {Suur-Uski}, {Tauber}, {Tavagnacco}, {Tenti}, {Toffolatti},
  {Tomasi}, {Trombetti}, {Valenziano}, {Valiviita}, {Van Tent}, {Vibert},
  {Vielva}, {Villa}, {Vittorio}, {Wandelt}, {Wehus}, {White}, {White},
  {Zacchei}, \& {Zonca}}]{Planck2018_cosmo}
{Planck Collaboration}, {Aghanim}, N., {Akrami}, Y., {et~al.}
  2020{\natexlab{b}}, \aap, 641, A6

\bibitem[{{Qin} {et~al.}(2020){Qin}, {Poulin}, {Mesinger}, {Greig}, {Murray},
  \& {Park}}]{QinPoulin_2020}
{Qin}, Y., {Poulin}, V., {Mesinger}, A., {et~al.} 2020, \mnras, 499, 550

\bibitem[{{Reichardt} {et~al.}(2021){Reichardt}, {Patil}, {Ade}, {Anderson},
  {Austermann}, {Avva}, {Baxter}, {Beall}, {Bender}, {Benson}, {Bianchini},
  {Bleem}, {Carlstrom}, {Chang}, {Chaubal}, {Chiang}, {Chou}, {Citron},
  {Corbett Moran}, {Crawford}, {Crites}, {de Haan}, {Dobbs}, {Everett},
  {Gallicchio}, {George}, {Gilbert}, {Gupta}, {Halverson}, {Harrington},
  {Henning}, {Hilton}, {Holder}, {Holzapfel}, {Hrubes}, {Huang}, {Hubmayr},
  {Irwin}, {Knox}, {Lee}, {Li}, {Lowitz}, {Luong-Van}, {McMahon}, {Mehl},
  {Meyer}, {Millea}, {Mocanu}, {Mohr}, {Montgomery}, {Nadolski}, {Natoli},
  {Nibarger}, {Noble}, {Novosad}, {Omori}, {Padin}, {Pryke}, {Ruhl},
  {Saliwanchik}, {Sayre}, {Schaffer}, {Shirokoff}, {Sievers}, {Smecher},
  {Spieler}, {Staniszewski}, {Stark}, {Tucker}, {Vand erlinde}, {Veach},
  {Vieira}, {Wang}, {Whitehorn}, {Williamson}, {Wu}, \&
  {Yefremenko}}]{Reichardt2020}
{Reichardt}, C.~L., {Patil}, S., {Ade}, P.~A.~R., {et~al.} 2021, \apj, 908, 199

\bibitem[{{Reichardt} {et~al.}(2012){Reichardt}, {Shaw}, {Zahn}, {Aird},
  {Benson}, {Bleem}, {Carlstrom}, {Chang}, {Cho}, {Crawford}, {Crites}, {de
  Haan}, {Dobbs}, {Dudley}, {George}, {Halverson}, {Holder}, {Holzapfel},
  {Hoover}, {Hou}, {Hrubes}, {Joy}, {Keisler}, {Knox}, {Lee}, {Leitch},
  {Lueker}, {Luong-Van}, {McMahon}, {Mehl}, {Meyer}, {Millea}, {Mohr},
  {Montroy}, {Natoli}, {Padin}, {Plagge}, {Pryke}, {Ruhl}, {Schaffer},
  {Shirokoff}, {Spieler}, {Staniszewski}, {Stark}, {Story}, {van Engelen},
  {Vand erlinde}, {Vieira}, \& {Williamson}}]{reichardt_2012_spt}
{Reichardt}, C.~L., {Shaw}, L., {Zahn}, O., {et~al.} 2012, \apj, 755, 70

\bibitem[{{Robertson} {et~al.}(2015){Robertson}, {Ellis}, {Furlanetto}, \&
  {Dunlop}}]{Robertson2015}
{Robertson}, B.~E., {Ellis}, R.~S., {Furlanetto}, S.~R., \& {Dunlop}, J.~S.
  2015, ApJ Letters, 802, L19

\bibitem[{{Roy} {et~al.}(2021){Roy}, {Kulkarni}, {Meerburg}, {Challinor},
  {Baccigalupi}, {Lapi}, \& {Haehnelt}}]{RoyKulkarni_2021}
{Roy}, A., {Kulkarni}, G., {Meerburg}, P.~D., {et~al.} 2021, \jcap, 2021, 003

\bibitem[{{Ruhl} {et~al.}(2004){Ruhl}, {Ade}, {Carlstrom}, {Cho}, {Crawford},
  {Dobbs}, {Greer}, {Halverson}, {Holzapfel}, {Lanting}, {Lee}, {Leitch},
  {Leong}, {Lu}, {Lueker}, {Mehl}, {Meyer}, {Mohr}, {Padin}, {Plagge}, {Pryke},
  {Runyan}, {Schwan}, {Sharp}, {Spieler}, {Staniszewski}, \& {Stark}}]{SPT}
{Ruhl}, J., {Ade}, P. A.~R., {Carlstrom}, J.~E., {et~al.} 2004, in Society of
  Photo-Optical Instrumentation Engineers (SPIE) Conference Series, Vol. 5498,
  Z-Spec: a broadband millimeter-wave grating spectrometer: design,
  construction, and first cryogenic measurements, ed. C.~M. {Bradford},
  P.~A.~R. {Ade}, J.~E. {Aguirre}, J.~J. {Bock}, M.~{Dragovan}, L.~{Duband},
  L.~{Earle}, J.~{Glenn}, H.~{Matsuhara}, B.~J. {Naylor}, H.~T. {Nguyen},
  M.~{Yun}, \& J.~{Zmuidzinas}, 11--29

\bibitem[{{Salvati} {et~al.}(2018){Salvati}, {Douspis}, \&
  {Aghanim}}]{Salvati2018}
{Salvati}, L., {Douspis}, M., \& {Aghanim}, N. 2018, \aap, 614, A13

\bibitem[{{Salvati} {et~al.}(2019){Salvati}, {Douspis}, {Ritz}, {Aghanim}, \&
  {Babul}}]{Salvati2019}
{Salvati}, L., {Douspis}, M., {Ritz}, A., {Aghanim}, N., \& {Babul}, A. 2019,
  \aap, 626, A27

\bibitem[{{Sato-Polito} {et~al.}(2021){Sato-Polito}, {Bernal}, {Boddy}, \&
  {Kamionkowski}}]{Sato-PolitoBernal_2021}
{Sato-Polito}, G., {Bernal}, J.~L., {Boddy}, K.~K., \& {Kamionkowski}, M. 2021,
  \prd, 103, 083519

\bibitem[{{Schenker} {et~al.}(2013){Schenker}, {Robertson}, {Ellis}, {Ono},
  {McLure}, {Dunlop}, {Koekemoer}, {Bowler}, {Ouchi}, {Curtis-Lake}, {Rogers},
  {Schneider}, {Charlot}, {Stark}, {Furlanetto}, \&
  {Cirasuolo}}]{schenker_2013}
{Schenker}, M.~A., {Robertson}, B.~E., {Ellis}, R.~S., {et~al.} 2013, \apj,
  768, 196

\bibitem[{{Seiler} {et~al.}(2019){Seiler}, {Hutter}, {Sinha}, \&
  {Croton}}]{rsage}
{Seiler}, J., {Hutter}, A., {Sinha}, M., \& {Croton}, D. 2019, \mnras, 1578

\bibitem[{{Shang} {et~al.}(2012){Shang}, {Haiman}, {Knox}, \&
  {Oh}}]{ShangHaiman_2012}
{Shang}, C., {Haiman}, Z., {Knox}, L., \& {Oh}, S.~P. 2012, \mnras, 421, 2832

\bibitem[{{Shaw} {et~al.}(2010){Shaw}, {Nagai}, {Bhattacharya}, \&
  {Lau}}]{ShawNagai_2010}
{Shaw}, L.~D., {Nagai}, D., {Bhattacharya}, S., \& {Lau}, E.~T. 2010, \apj,
  725, 1452

\bibitem[{{Shaw} {et~al.}(2012){Shaw}, {Rudd}, \& {Nagai}}]{ShawRudd_2012}
{Shaw}, L.~D., {Rudd}, D.~H., \& {Nagai}, D. 2012, \apj, 756, 15

\bibitem[{{Sievers} {et~al.}(2013){Sievers}, {Hlozek}, {Nolta}, {Acquaviva},
  {Addison}, {Ade}, {Aguirre}, {Amiri}, {Appel}, {Barrientos}, {Battistelli},
  {Battaglia}, {Bond}, {Brown}, {Burger}, {Calabrese}, {Chervenak}, {Crichton},
  {Das}, {Devlin}, {Dicker}, {Bertrand Doriese}, {Dunkley}, {D{\"u}nner},
  {Essinger-Hileman}, {Faber}, {Fisher}, {Fowler}, {Gallardo}, {Gordon},
  {Gralla}, {Hajian}, {Halpern}, {Hasselfield}, {Hern{\'a}ndez-Monteagudo},
  {Hill}, {Hilton}, {Hilton}, {Hincks}, {Holtz}, {Huffenberger}, {Hughes},
  {Hughes}, {Infante}, {Irwin}, {Jacobson}, {Johnstone}, {Baptiste Juin},
  {Kaul}, {Klein}, {Kosowsky}, {Lau}, {Limon}, {Lin}, {Louis}, {Lupton},
  {Marriage}, {Marsden}, {Martocci}, {Mauskopf}, {McLaren}, {Menanteau},
  {Moodley}, {Moseley}, {Netterfield}, {Niemack}, {Page}, {Page}, {Parker},
  {Partridge}, {Plimpton}, {Quintana}, {Reese}, {Reid}, {Rojas}, {Sehgal},
  {Sherwin}, {Schmitt}, {Spergel}, {Staggs}, {Stryzak}, {Swetz}, {Switzer},
  {Thornton}, {Trac}, {Tucker}, {Uehara}, {Visnjic}, {Warne}, {Wilson},
  {Wollack}, {Zhao}, \& {Zunckel}}]{SieversHlozek_2013}
{Sievers}, J.~L., {Hlozek}, R.~A., {Nolta}, M.~R., {et~al.} 2013, \jcap, 2013,
  060

\bibitem[{{Sobrin} {et~al.}(2022){Sobrin}, {Anderson}, {Bender}, {Benson},
  {Dutcher}, {Foster}, {Goeckner-Wald}, {Montgomery}, {Nadolski}, {Rahlin},
  {Ade}, {Ahmed}, {Anderes}, {Archipley}, {Austermann}, {Avva}, {Aylor},
  {Balkenhol}, {Barry}, {Thakur}, {Benabed}, {Bianchini}, {Bleem}, {Bouchet},
  {Bryant}, {Byrum}, {Carlstrom}, {Carter}, {Cecil}, {Chang}, {Chaubal},
  {Chen}, {Cho}, {Chou}, {Cliche}, {Crawford}, {Cukierman}, {Daley}, {Haan},
  {Denison}, {Dibert}, {Ding}, {Dobbs}, {Everett}, {Feng}, {Ferguson}, {Fu},
  {Galli}, {Gambrel}, {Gardner}, {Gualtieri}, {Guns}, {Gupta}, {Guyser},
  {Halverson}, {Harke-Hosemann}, {Harrington}, {Henning}, {Hilton}, {Hivon},
  {Holder}, {Holzapfel}, {Hood}, {Howe}, {Huang}, {Irwin}, {Jeong}, {Jonas},
  {Jones}, {Khaire}, {Knox}, {Kofman}, {Korman}, {Kubik}, {Kuhlmann}, {Kuo},
  {Lee}, {Leitch}, {Lowitz}, {Lu}, {Meyer}, {Michalik}, {Millea}, {Natoli},
  {Nguyen}, {Noble}, {Novosad}, {Omori}, {Padin}, {Pan}, {Paschos}, {Pearson},
  {Posada}, {Prabhu}, {Quan}, {Reichardt}, {Riebel}, {Riedel}, {Rouble},
  {Ruhl}, {Saliwanchik}, {Sayre}, {Schiappucci}, {Shirokoff}, {Smecher},
  {Stark}, {Stephen}, {Story}, {Suzuki}, {Tandoi}, {Thompson}, {Thorne},
  {Tucker}, {Umilta}, {Vale}, {Vanderlinde}, {Vieira}, {Wang}, {Whitehorn},
  {Wu}, {Yefremenko}, {Yoon}, \& {Young}}]{SPT3G}
{Sobrin}, J.~A., {Anderson}, A.~J., {Bender}, A.~N., {et~al.} 2022, \apjs, 258,
  42

\bibitem[{{Sunyaev} \& {Zeldovich}(1980)}]{sunyaev_zeldovich_1980}
{Sunyaev}, R.~A. \& {Zeldovich}, I.~B. 1980, \araa, 18, 537

\bibitem[{{Taburet} {et~al.}(2009){Taburet}, {Aghanim}, {Douspis}, \&
  {Langer}}]{2009MNRAS.392.1153T}
{Taburet}, N., {Aghanim}, N., {Douspis}, M., \& {Langer}, M. 2009, \mnras, 392,
  1153

\bibitem[{{Tanimura} {et~al.}(2022){Tanimura}, {Douspis}, {Aghanim}, \&
  {Salvati}}]{TanimuraDouspis_2021}
{Tanimura}, H., {Douspis}, M., {Aghanim}, N., \& {Salvati}, L. 2022, \mnras,
  509, 300

\bibitem[{{Tashiro} {et~al.}(2011){Tashiro}, {Aghanim}, {Langer}, {Douspis},
  {Zaroubi}, \& {Jeli{\'c}}}]{TashiroAghanim_2011}
{Tashiro}, H., {Aghanim}, N., {Langer}, M., {et~al.} 2011, \mnras, 414, 3424

\bibitem[{{Trac} {et~al.}(2022){Trac}, {Chen}, {Holst}, {Alvarez}, \&
  {Cen}}]{TracChen_2022}
{Trac}, H., {Chen}, N., {Holst}, I., {Alvarez}, M.~A., \& {Cen}, R. 2022, \apj,
  927, 186

\bibitem[{{Viero} {et~al.}(2013){Viero}, {Wang}, {Zemcov}, {Addison},
  {Amblard}, {Arumugam}, {Aussel}, {B{\'e}thermin}, {Bock}, {Boselli}, {Buat},
  {Burgarella}, {Casey}, {Clements}, {Conley}, {Conversi}, {Cooray}, {De
  Zotti}, {Dowell}, {Farrah}, {Franceschini}, {Glenn}, {Griffin},
  {Hatziminaoglou}, {Heinis}, {Ibar}, {Ivison}, {Lagache}, {Levenson},
  {Marchetti}, {Marsden}, {Nguyen}, {O'Halloran}, {Oliver}, {Omont}, {Page},
  {Papageorgiou}, {Pearson}, {P{\'e}rez-Fournon}, {Pohlen}, {Rigopoulou},
  {Roseboom}, {Rowan-Robinson}, {Schulz}, {Scott}, {Seymour}, {Shupe}, {Smith},
  {Symeonidis}, {Vaccari}, {Valtchanov}, {Vieira}, {Wardlow}, \&
  {Xu}}]{VieroWang_2013}
{Viero}, M.~P., {Wang}, L., {Zemcov}, M., {et~al.} 2013, \apj, 772, 77

\bibitem[{{Zahn} {et~al.}(2007){Zahn}, {Lidz}, {McQuinn}, {Dutta}, {Hernquist},
  {Zaldarriaga}, \& {Furlanetto}}]{ZahnLidz_2007}
{Zahn}, O., {Lidz}, A., {McQuinn}, M., {et~al.} 2007, \apj, 654, 12

\bibitem[{{Zahn} {et~al.}(2011){Zahn}, {Mesinger}, {McQuinn}, {Trac}, {Cen}, \&
  {Hernquist}}]{ZahnMesinger_2011}
{Zahn}, O., {Mesinger}, A., {McQuinn}, M., {et~al.} 2011, \mnras, 414, 727

\bibitem[{{Zahn} {et~al.}(2012){Zahn}, {Reichardt}, {Shaw}, {Lidz}, {Aird},
  {Benson}, {Bleem}, {Carlstrom}, {Chang}, {Cho}, {Crawford}, {Crites}, {de
  Haan}, {Dobbs}, {Dor{\'e}}, {Dudley}, {George}, {Halverson}, {Holder},
  {Holzapfel}, {Hoover}, {Hou}, {Hrubes}, {Joy}, {Keisler}, {Knox}, {Lee},
  {Leitch}, {Lueker}, {Luong-Van}, {McMahon}, {Mehl}, {Meyer}, {Millea},
  {Mohr}, {Montroy}, {Natoli}, {Padin}, {Plagge}, {Pryke}, {Ruhl}, {Schaffer},
  {Shirokoff}, {Spieler}, {Staniszewski}, {Stark}, {Story}, {van Engelen},
  {Vanderlinde}, {Vieira}, \& {Williamson}}]{zahn_2012_spt}
{Zahn}, O., {Reichardt}, C.~L., {Shaw}, L., {et~al.} 2012, \apj, 756, 65

\bibitem[{Zeldovich \& Sunyaev(1969)}]{zeldovich_sunyaev_1969}
Zeldovich, Y.~B. \& Sunyaev, R.~A. 1969, Astrophysics and Space Science, 4, 301

\end{thebibliography}

\appendix
\section{Random forests predictions}
\label{app:ML}

The derivation of the kSZ angular power spectrum introduced in \citet{Gorce2020} is too time-consuming to be included as such in a sampling algorithm. For this reason, and as already successfully implemented in \citetalias{PAPER1} to estimate the tSZ power, we train random forests to approximate the kSZ angular power spectrum, given a set of cosmological and reionisation parameters. We refer the reader to \citetalias{PAPER1} for more details about our choice of machine learning technique and parameter optimisation.

We use for training a set of $50\,000$ kSZ power spectra computed on 25 $\ell$-bins ($100 \leq \ell \leq 10\,500$)\footnote{Given the smoothness of the evolution of the kSZ spectra with $\ell$, we find this number of bins to be sufficient.} for parameters $\mathrm{ln}(10^{10}~A_s)$, $\Omega_{\mathrm{b}}h^2$, $\Omega_{\mathrm{c}}h^2$, $100~\theta_\mathrm{MC}$, $n_s$, $\zre$, $\mathrm{d}z$, $\log \alpha_0$ and $\kappa$. The parameter ranges covered by the training set include the sampling ranges of Sec.~\ref{subsec:data_spt}. The samples are divided into fifths such that 60\%, 20\% and 20\% of the models are used for the learning, validation, and testing phase, respectively. Because of their differences in shape and amplitude, as well as their sensitivity to different cosmological parameters, we train two separate sets of random forests to predict the late-time and patchy kSZ signals.

\begin{figure}
    \centering
    \includegraphics[width=\columnwidth]{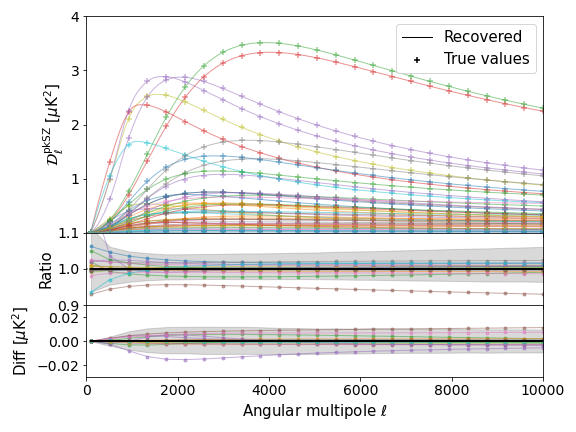}\\
    \includegraphics[width=\columnwidth]{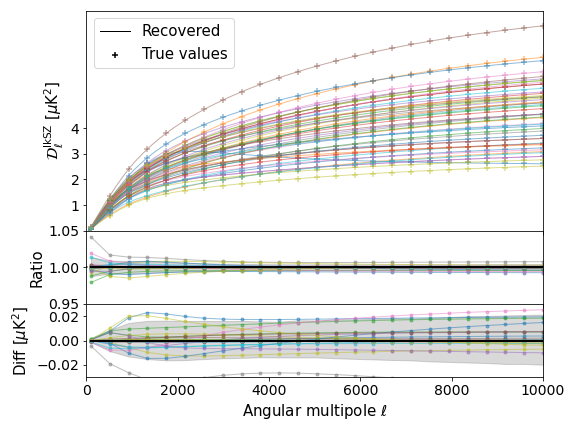}
    \caption{Patchy (upper panel) and late-time (low panel) kSZ spectra recovered by our random forest regressors, compared to true values. The lower panels show the $95\%$ confidence intervals on the ratio of the two and the  $95\%$ confidence intervals on the absolute difference between the two, that is the absolute error.}
    \label{fig:ksz_spectra_training}
\end{figure}

\begin{figure*}
    \centering
    \includegraphics[width=.95\textwidth]{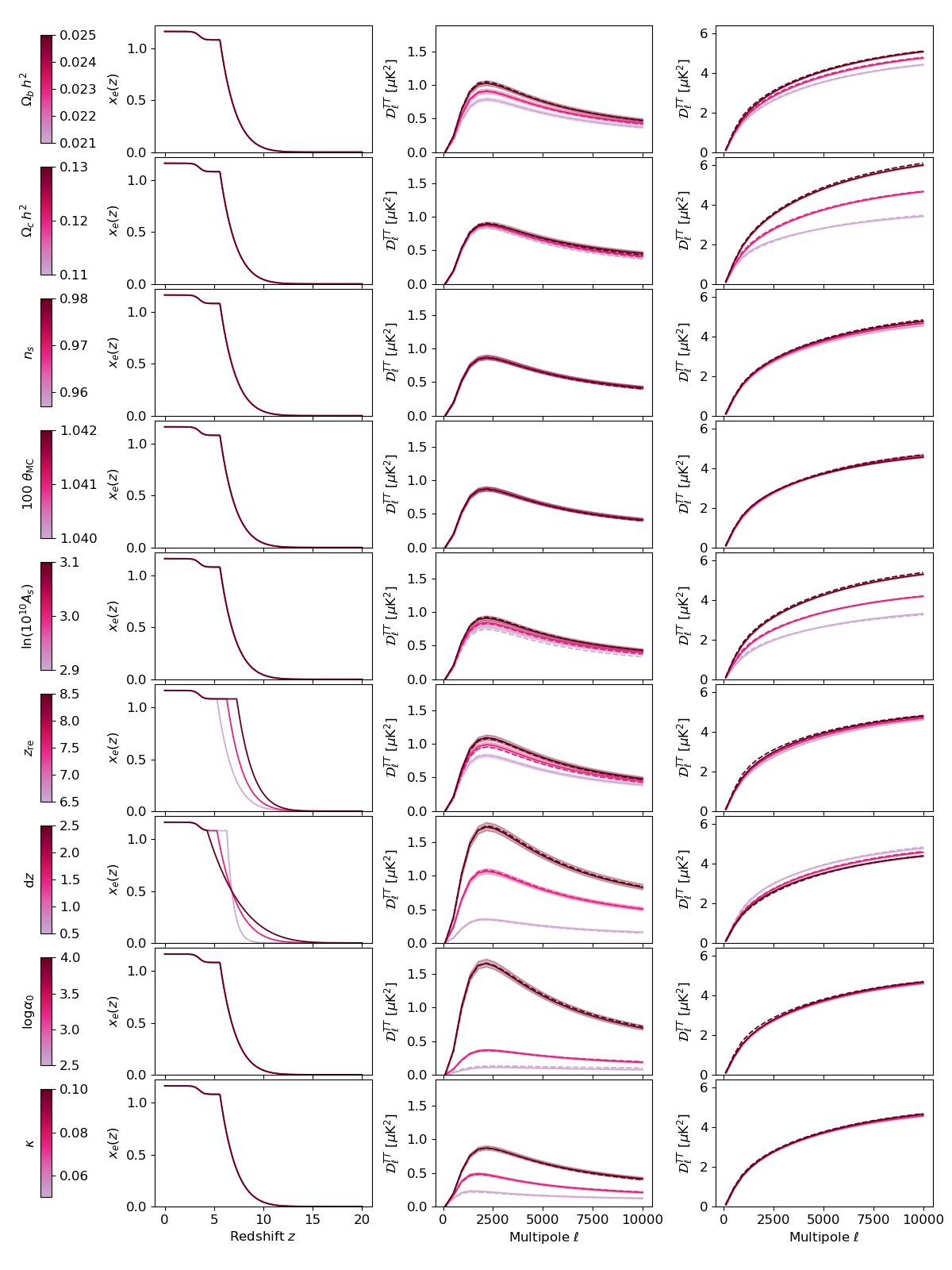}
    \caption{Dependence of the reionisation history (left panels), patchy (middle panels) and late-time (right panels) kSZ angular power spectra on the cosmological parameters the random forests were trained on (rows). For each parameter, three values are tested and we compare the spectrum obtained with the exact computation (dashed line) from the one inferred by the RF (solid line). The shaded area corresponds to the $68\%$ confidence interval corresponding to the RF reconstruction errors (identical as Fig.~\ref{fig:ksz_spectra_training}).}
    \label{fig:dep_parameters}
\end{figure*}

In order to facilitate the training, we scale each spectrum in the training set by $\mathcal{D}_{3000}$ values obtained with the method described below. Doing so, the amplitude differences between the spectra in the training set will be smoothed out, and the random forests will focus their training on the dependency between parameters and the shape of the spectrum, rather than its amplitude. We use an MCMC sampling algorithm to obtain the parameters $\{ \alpha_i \}$ that minimise the distance between the true $\mathcal{D}_{3000}^{\rm true}$ and a model given by $\mathcal{D}_{3000}^{\rm model} \equiv A_0 \prod_i \left[\theta_i/\theta_{\mathrm{ref},i}\right] ^{\alpha_i}$, for each of the cosmological and reionisation parameter sets $\{ \theta_i\}$ in the training set. We fix $A_0$ to the amplitude of the spectrum at $\ell=3000$ obtained with the full computation for $\{\theta_{\mathrm{ref},i}\}$. 
We obtain the following for for the patchy kSZ amplitude:
\begin{equation}
\begin{aligned}
    \mathcal{D}_{3000}^\mathrm{pkSZ} = 1.03~\mu\mathrm{K}^2 &\times  \left( \frac{\zre}{7.36} \right)^{0.70} \left( \frac{\mathrm{d}z}{1.4} \right)^{0.87} \left( \frac{\log \alpha_0}{3.7} \right)^{6.60} \left( \frac{\kappa}{0.10} \right)^{1.99}  \\
    & \times \left( \frac{\Omega_b h^2}{0.0224} \right)^{1.47} \left( \frac{\Omega_c h^2}{0.120} \right)^{0.20} \left( \frac{n_s}{0.9653} \right)^{-0.28} \\ &\times \left( \frac{100\theta_\mathrm{MC}}{1.041} \right)^{0.012} \left( \frac{\log A}{3.048} \right)^{2.18},
\end{aligned}
\end{equation}
and the following for the late-time kSZ amplitude:
\begin{equation}
\begin{aligned}
    \mathcal{D}_{3000}^\mathrm{lkSZ} = 3.11~\mu\mathrm{K}^2 &\times  \left( \frac{\zre}{7.37} \right)^{0.24} \left( \frac{\mathrm{d}z}{1.4} \right)^{-0.08} \left( \frac{\log \alpha_0}{3.7} \right)^{0.07} \left( \frac{\kappa}{0.10} \right)^{0.04}  \\
    & \times \left( \frac{\Omega_b h^2}{0.0224} \right)^{0.78} \left( \frac{\Omega_c h^2}{0.120} \right)^{2.93} \left( \frac{n_s}{0.9653} \right)^{2.12} \\ &\times \left( \frac{100\theta_\mathrm{MC}}{1.041} \right)^{-0.006} \left( \frac{\log A}{3.048} \right)^{6.94}.
\end{aligned}
\end{equation}
We note that the observed scalings, although model-dependent, exhibit the expected dependencies. For example, the amplitude of the late-time kSZ is inversely proportional to the duration of reionisation, $\mathrm{d}z$, but proportional to the reionisation redshift $\zre$. Indeed, for a given $\zre$, a longer reionisation will end later, and, therefore, the late-time signal will be integrated over a short redshift range. Conversely, for a given reionisation duration, a larger midpoint means that reionisation will end earlier and the late-time signal will be integrated over a larger redshift range. Additionally, the strong dependency of both kSZ components to the baryon and cold dark matter densities is related to the large contribution of the matter power spectrum in the derivation of their spectra. These dependencies are illustrated on Fig.~\ref{fig:dep_parameters} and we refer the interested reader to \citet{ChenTrac_2022} for a detailed discussion of the dependency of the shape and amplitude of the patchy kSZ on various reionisation parameters. We emphasise the fact that the amplitude of the patchy and late-time kSZ spectra increases slowly with $\zre$, which will limit the constraining power of kSZ measurements on this parameter.

\begin{figure*}
    \centering
    \includegraphics[width=\textwidth]{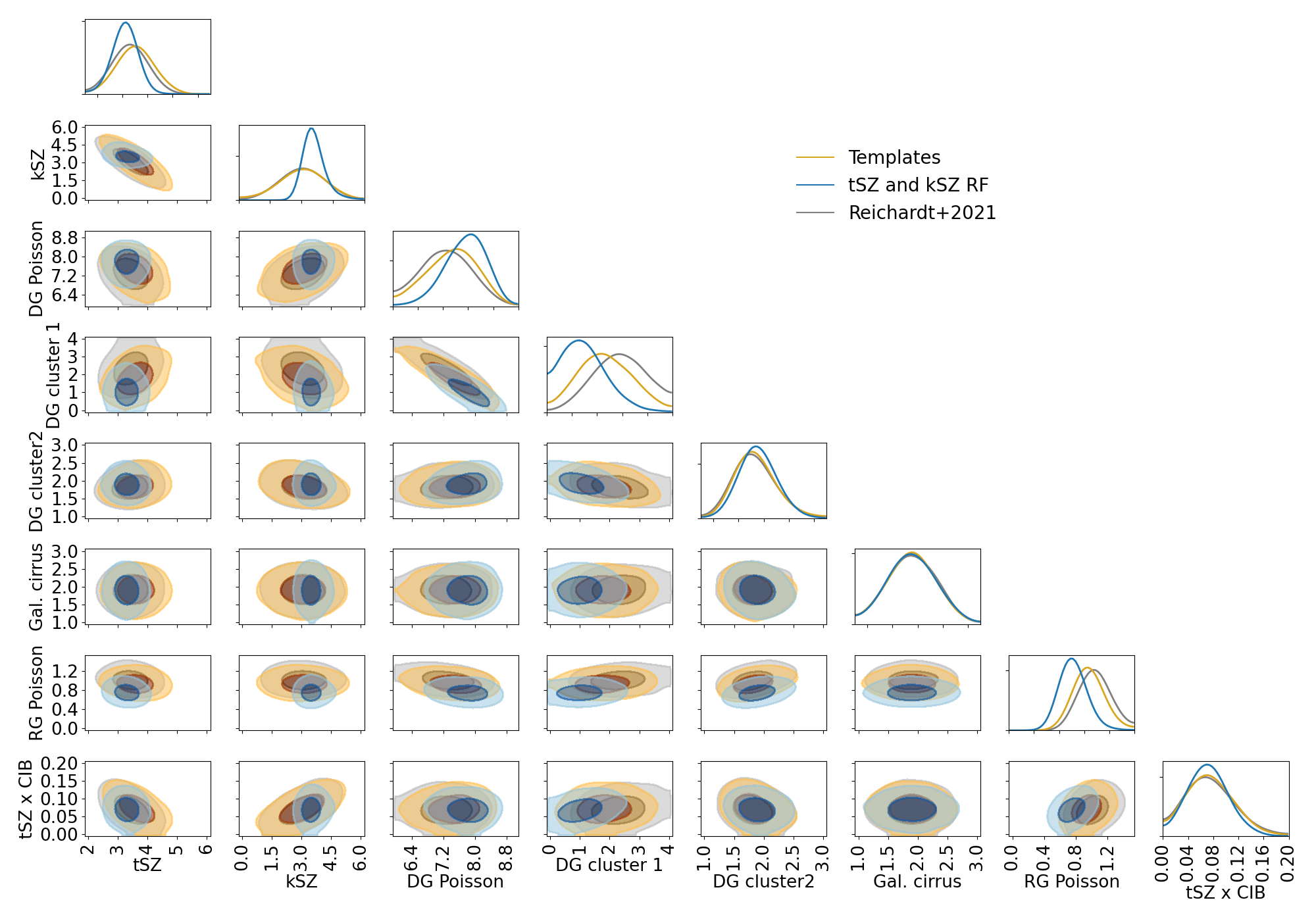}
    \caption{Posterior probability distributions of nuisance parameters obtained with SPT high-$\ell$ data \citepalias{Reichardt2020} using either templates (in yellow) or analytic derivation (in blue) for the tSZ and the kSZ power and a physical parameterisation of reionisation. Inferences are compared to the initial analysis, with templates and fixed reionisation (in grey). The CIBxtSZ amplitude is given in terms of its absolute value.}
    \label{fig:app_nuisance_reichardt_comparison}
\end{figure*}

We show in Fig.~\ref{fig:ksz_spectra_training} some of the kSZ spectra recovered by the random forests after training, using parameters in the testing set -- that is, parameter sets that the RF has not been explicitly trained on. We see that, overall, the late-time regressor performs better than the patchy one. They obtain, respectively, scores of 0.993 and 0.998 for a maximum of 1.0, corresponding to the coefficient of determination $R^2$ of the prediction. This is likely because, as can be seen on the figure, there is much more variance between the patchy spectra than between the late-time spectra, and in particular there is more variance in their shape. However, in both cases, reconstruction errors remain within $5\%$ of the true value ($1\%$ for late-time), and are recovered with an absolute error $<0.02~\mu\mathrm{K}^2$, which is about ten times smaller than the precision of the constraints obtained with the data (see Table~\ref{tab:results}). These errors are constant on the range of multipoles covered by the SPT data. To optimise the quality of the prediction, we have tested different values of hyper-parameters used by the RF, and we choose 40 for the number of trees and a depth of 20. These values maximise the score, whilst limiting the time and memory imprints of the reconstruction. 


\section{Full parameter distributions}

We present in this appendix additional posterior distributions from the different analyses carried out in this work, in order to support our arguments.

\subsection{Fixed cosmology}

We present in this appendix the posterior  distributions of the nuisance parameters -- that is, of the amplitudes at $\ell=3000$ of the various foregrounds considered, for the run fitting the SPT data with fixed cosmology. Results are compared in Fig.~\ref{fig:app_nuisance_reichardt_comparison} between the results of \citetalias{Reichardt2020}, using templates for both SZ spectra; a similar analysis but with the reionisation model of Eq.~\eqref{eq:xe_param}; or with RF-inferred SZ spectra (see above). The constraints obtained with the cosmology-dependent SZ spectra are fully consistent with previous results, although tighter. For example, the error bars on the contribution from Poisson-distributed dusty galaxies are reduced by $15\%$. Overall, the contribution of dusty and radio galaxies to the observed temperature power spectrum is smaller on all scales.
Regarding the tSZxCIB correlations, the constraints on their amplitude at $\ell=3000$, $\xi$, obtained with the RF are fully consistent with previous results - that is, the data favours positive correlations at the $2\sigma$ level, but the new model breaks the degeneracy between $\xi$ and the SZ amplitudes.

\begin{figure*}
    \centering
    \includegraphics[width=\textwidth]{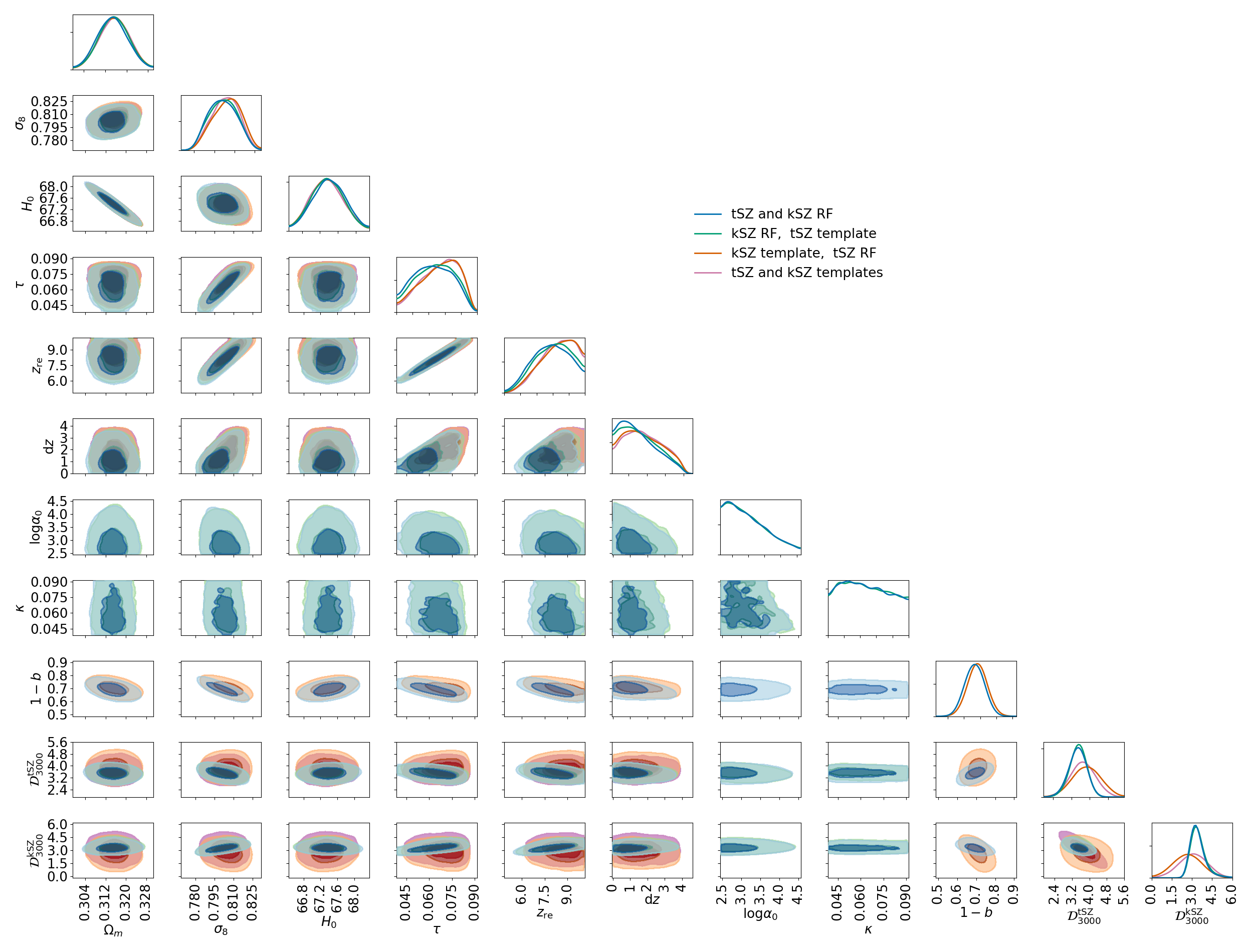}
    \caption{Posterior distributions of the different parameters sampled in the run with free cosmological and reionisation parameters using RF-predicted kSZ and tSZ spectra.}
    \label{fig:full_triangle_plot}
\end{figure*}

\subsection{Varying cosmology}

\begin{figure*}
    \centering
    \includegraphics[width=.8\textwidth]{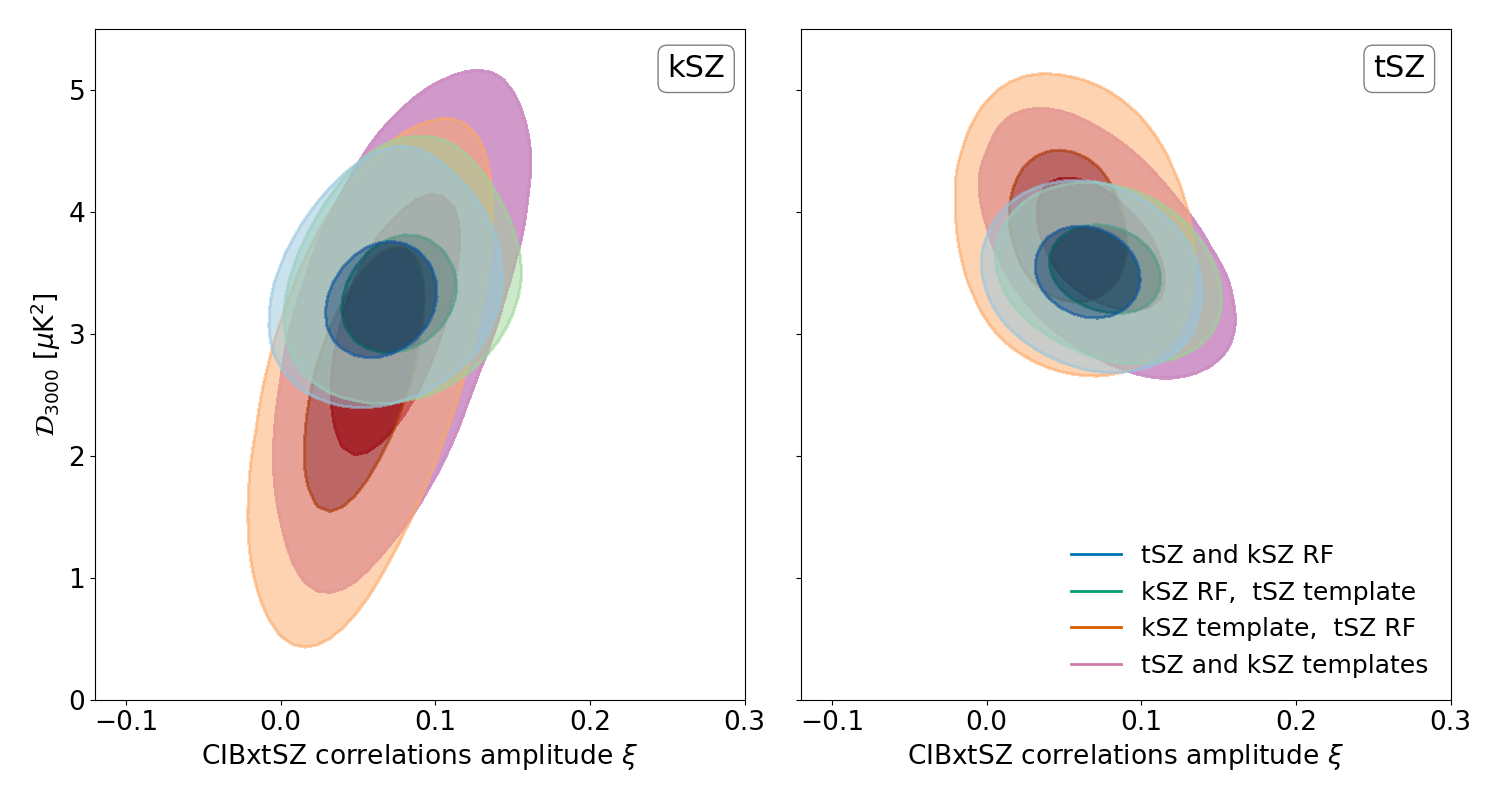}
    \caption{Joint posterior distributions of the kSZ and tSZ amplitudes at $\ell=3000$ with the amplitude of the tSZxCIB correlations at the same multipole, $\xi$. In all the results presented, the cosmology is free to vary, and different methods are used to model the contribution of the SZ effects to the observed small-scale temperature power spectrum of the CMB.}
    \label{fig:tSZxCIB_free_cosmo}
\end{figure*}

In Fig.~\ref{fig:full_triangle_plot}, we present the posterior distribution of the main cosmological and nuisance parameters obtained in Sec.~\ref{subsec:res_small_scale}. Depending on the nature of the analysis (that is, if templates or RF are used), these parameters can be sampled or derived. The cosmological information included in the SZ spectra when they are inferred by the RF at each iteration of the chain favours later and quicker reionisation scenarios although the constraints remain fully consistent with the results obtained with templates. The anti-correlation between $\sigma_8$ and $1-b$, as well as the correlation between $1-b$ and the tSZ amplitude, already observed in \citetalias{PAPER1}, is not broken by the addition of the cosmology-dependent kSZ model.
Finally, the kSZ shape parameters, $\log \alpha_0$ and $\kappa$, cannot be constrained precisely. Only an upper limit $\log [ \alpha_0/\mathrm{Mpc}^3] < 4.1$ ($95\%$ confidence) can be placed on the amplitude of the large-scale amplitude of the free electrons density fluctuations at high redshift. The interpretation of such an upper limit, in terms of high-redshift galaxy properties, will be the focus of future work.

We show in Fig.~\ref{fig:tSZxCIB_free_cosmo} the joint posterior distributions of the amount of CIBxtSZ correlations, $\xi$, with the amplitudes of the kSZ and tSZ power spectra at $\ell=3000$, for the four cases aforementioned. We see that, despite being fully consistent with the results obtained with templates, using the cosmology-dependent SZ spectra allows to break the degeneracy between $\xi$ and the SZ amplitudes. Replacing the tSZ template by the RF-inferred spectrum in the analysis allows to break the degeneracy between $\xi$ and the tSZ amplitude: the correlation coefficient between $\DtSZ$ and $\xi$ drops from $- 0.6$ when the model used for the tSZ is a template, to $- 0.2$ when the cosmology-dependent tSZ power is used. Similarly, exploiting the cosmological information included in the kSZ power breaks the degeneracy between $\DkSZ$ and $\xi$, moving from $0.6$ to $0.2$ when the kSZ template is replaced by the RF.

\end{document}